# Streaming Symmetric Norms via Measure Concentration[*]

Jarosław Błasiok[†]   Vladimir Braverman[‡]   Stephen R. Chestnut[§]

Robert Krauthgamer[¶]   Lin F. Yang[‖]

June 9, 2017

**Abstract**

We characterize the streaming space complexity of every symmetric norm $l$ (a norm on $\mathbb{R}^n$ invariant under sign-flips and coordinate-permutations), by relating this space complexity to the measure-concentration characteristics of $l$. Specifically, we provide nearly matching upper and lower bounds on the space complexity of calculating a $(1 \pm \epsilon)$-approximation to the norm of the stream, for every $0 < \epsilon \le 1/2$. (The bounds match up to $\operatorname{poly}(\epsilon^{-1} \log n)$ factors.) We further extend those bounds to any large approximation ratio $D \ge 1.1$, showing that the decrease in space complexity is proportional to $D^2$, and that this factor the best possible. All of the bounds depend on the median of $l(x)$ when $x$ is drawn uniformly from the $l_2$ unit sphere. The same median governs many phenomena in high-dimensional spaces, such as large-deviation bounds and the critical dimension in Dvoretzky's Theorem.

The family of symmetric norms contains several well-studied norms, such as all $l_p$ norms, and indeed we provide a new explanation for the disparity in space complexity between $p \le 2$ and $p > 2$. In addition, we apply our general results to easily derive bounds for several norms that were not studied before in the streaming model, including the top-$k$ norm and the $k$-support norm, which was recently employed for machine learning tasks.

Overall, these results make progress on two outstanding problems in the area of sublinear algorithms (Problems 5 and 30 in `http://sublinear.info`).

## 1 Introduction

The study of norms on data streams has a rich history, and in particular has driven much of the fantastic development of streaming algorithms, see e.g. [AMS99, IW05, Ind06, Mut05]. A *data stream* is a sequence of additive $\pm 1$ updates that accumulate on the coordinates of an $n$-dimensional vector $v$, and a streaming algorithm reads the sequence of updates and computes some function of $v$. This is known as the turnstile model, and for simplicity we assume that $|v_i| \le \operatorname{poly}(n)$, for all $i \in [n]$. Despite plenty of work, it is still an open problem to design a generic streaming algorithm for approximating norms. Although very challenging, it may not be too much to ask for. In fact, several existing methods, including the Indyk-Woodruff sketch [IW05, BOR15], yield so-called "universal sketches" that can be used to approximate whole classes of streaming problems at once. So we ask, is there a *generic method* that can approximate any desired norm of a stream with near-optimal space complexity? Second, is there a *universal sketch* whose single evaluation on a vector (say on a stream) suffices to approximate every norm in a wide class? While several powerful upper and lower bound techniques have been developed, including embeddings, heavy-hitters, and reductions from Communication Complexity, it is not apparent how they can be applied to an

---

[*]A preliminary version of this paper was published in the proceedings of STOC 2017.

[†]Harvard University, Cambridge, MA, USA. `jblasiok@g.harvard.edu`. Supported by NSF grant CCF-1350670.

[‡]Johns Hopkins University, USA. Email: `vova@cs.jhu.edu`. This material is based upon work supported by the NSF Grants IIS-1447639, EAGER CCF- 1650041, and CAREER CCF-1652257.

[§]ETH Zurich, Switzerland. Email: `stephenc@ethz.ch`

[¶]Weizmann Institute of Science, Israel. Email: `robert.krauthgamer@weizmann.ac.il` Work supported in part by the Israel Science Foundation grant #897/13 and by a Minerva Foundation grant.

[‖]Johns Hopkins University, USA. Email: `lyang@jhu.edu` This material is based upon work supported by the NSF Grant IIS-1447639.

entirely new norm, see also Open Problems 5 (Sketchable Distances) and 30 (Universal Sketching) in the list [sub06].

This is a real challenge for at least two reasons. First, we lack a generic framework for embeddings. Even when it is possible to embed into an easy-to-handle space, a new embedding must be constructed and applied to the input stream for each norm. Second, current techniques, heavy-hitters included, have been confined to norms with additive structure. Nearly all of the norms considered so far decompose, on some level, into a sum of independent quantities, and this fact is heavily exploited in the design of algorithms and lower bounds. Examples include $l_p$ norms (see references in Section 1.3), the entropy norm [CDM06, CCM07, HNO08], and cascaded $l_p$ norms [JW09, Jay13]. Abandoning our reliance on additive decomposability has been a major bottleneck en route to a broader characterization of norms.

We overcome this barrier in the setting of symmetric norms, see e.g. [Bha97, Chapter IV]. A norm $l : \mathbb{R}^n \to \mathbb{R}$ is called *symmetric* if, for all $x \in \mathbb{R}^n$ and every $n \times n$ permutation matrix $P$, it satisfies $l(Px) = l(x)$ and also $l(|x|) = l(x)$, where $|x|$ is the coordinate-wise absolute value of $x$. It is a partial answer to the question above, as we design a generic algorithm for symmetric norms and it is based on a universal sketch. Specifically, for every $s > 0$, there is a single sketch of size $s \cdot \text{poly}(\log(n)/\epsilon)$, that yields a $(1 \pm \epsilon)$-approximation[1] for every symmetric norm whose streaming space complexity is at most $s$. In fact, we show that the streaming space complexity of a symmetric norm is determined by the norm's measure-concentration characteristics. To be precise, let $X \in \mathbb{R}^n$ be uniformly distributed on $S^{n-1}$, the $l_2$ unit sphere. The median of a symmetric norm $l$ is the (unique!) value $\mathrm{M}_l$ such that $\Pr[l(X) \geq \mathrm{M}_l] \geq 1/2$ and $\Pr[l(X) \leq \mathrm{M}_l] \geq 1/2$. Similarly, $\mathfrak{b}_l$ denotes the maximum value of $l(x)$ over $x \in S^{n-1}$. We call the ratio
$$\mathrm{mc}(l) := \mathfrak{b}_l / \mathrm{M}_l$$
the *modulus of concentration* of the norm $l$. Our results show that this modulus of concentration is crucial in determining the streaming space complexity of any symmetric norm. This quantity governs many phenomena in high-dimensional spaces, for example, it appears in large-deviation bounds and the critical dimension in Dvoretzky's Theorem is $n/\mathrm{mc}(l)^2$, see e.g. [MS86, KV07].

Symmetric norms clearly include the $l_p$ and entropy norms, and we present fresh examples with heretofore unknown streaming space complexity, like the top-$k$ norm, $Q$ norms, and $Q'$ norms, later on. Although matrix norms are generally not symmetric, our results immediately imply lower bounds for unitarily invariant matrix norms, for example the Ky Fan norms, by restricting attention to diagonal matrices.

One well-studied family of symmetric norms is that of $l_p$ norms on $\mathbb{R}^n$, defined as $l_p(x) := (\sum_{i=1}^n |x_i|^p)^{1/p}$. For $1 \leq p \leq 2$, the maximum value of $l_p(x)$ over $x \in S^{n-1}$ is $\mathfrak{b}_{l_p} = n^{1/p-1/2}$ and concentrates at $\mathrm{M}_{l_p} = \Theta(n^{1/p-1/2})$, so the modulus of concentration is $\mathrm{mc}(l_p) = O(1)$. For $p > 2$, the maximum is $\mathfrak{b}_{l_p} = 1$ but again concentrates at $\mathrm{M}_{l_p} = \Theta(n^{1/p-1/2})$, hence $\mathrm{mc}(l_p) = \Theta(n^{1/2-1/p})$. Recall that the streaming space complexity for a $(1 \pm 1/10)$-approximation of $l_p$ is $\Theta(\log n)$, when $p \leq 2$ [KNW10], and is $\Theta(n^{1-2/p} \log n)$ when $p > 2$ [LW13, Gan15] (the constant $1/10$ here is arbitrary). Thus for all values of $p \geq 1$, the space complexity of computing a $(1 \pm 1/10)$-approximation to $l_p$ is $\Theta(\mathrm{mc}(l_p)^2 \log n)$. Our main result recovers this fact up to a polylog $n$ factor.

But, the modulus of concentration cannot be the whole story for streaming algorithms. It expresses an average behavior of the norm on $\mathbb{R}^n$, and even if the norm is well-behaved on average, like $l_1$ for example, it is possible that a more difficult norm is concealed in a lower-dimensional subspace. One example of this is $l(x) := \max\{l_\infty(x), l_1(x)/\sqrt{n}\}$ on $\mathbb{R}^n$, which has $\mathrm{mc}(l) = O(1)$. However, when $x$ has fewer than $\sqrt{n}$ nonzero coordinates, $l(x) = l_\infty(x)$, which is just a lower-dimensional copy of $l_\infty$ and implies, by [AMS99], an $\Omega(\sqrt{n})$ space lower bound for $l$. In order for the modulus of concentration to have any connection with streaming space complexity, we have to close this gap.

Notice that, for every $k \leq n$, the norm $l$ induces a norm $l^{(k)}$ on $\mathbb{R}^k$ by setting
$$l^{(k)}((x_1, x_2, \ldots, x_k)) := l((x_1, \ldots, x_k, 0, \ldots, 0)).$$
Of course, because of the permutation symmetry we could have chosen any set of $n - k$ coordinates to be the zeros. As the examples above show, the modulus of concentration of $l^{(k)}$ may vary with $k$. However, any streaming approximation algorithm for $l$ is also trivially a streaming approximation algorithm for $l^{(k)}$. We

---

[1] We state the approximation ratio in one of two standard ways. A $D$-approximation, $D \geq 1$, to $l(v)$ is a value $\hat{l}$ such that $l(v) \leq \hat{l} \leq Dl(v)$. When $D$ is very close to one, it is more convenient to consider a $(1 \pm \epsilon)$-approximation, $0 \leq \epsilon < 1/2$, which is defined as $(1 - \epsilon)l(v) \leq \hat{l} \leq (1 + \epsilon)l(v)$ and corresponds to a $D$-approximation for $D = \frac{1+\epsilon}{1-\epsilon}$.

therefore define the *maximum modulus of concentration* of the norm $l$ as
$$\mathrm{mmc}(l) := \max_{k \leq n} \mathrm{mc}(l^{(k)}) = \max_{k \leq n} \frac{\mathfrak{b}_{l^{(k)}}}{\mathrm{M}_{l^{(k)}}}.$$
Our main result is that this quantity characterizes the streaming space complexity of every symmetric norm $l$.

## 1.1 Our Results

Quite surprisingly, for every symmetric norm $l$ on $\mathbb{R}^n$, the optimal space complexity of a streaming algorithm that gives a $(1 \pm \epsilon)$-approximation for $l$ is $\mathrm{mmc}(l)^2 \cdot \mathrm{poly}(\log(n)/\epsilon)$. This characterization tells us in particular whether a given symmetric norm admits a polylogarithmic space approximation or requires polynomial space.

**Theorem 1.1** (Main Theorem)**.** *Let $l$ be a symmetric norm on $\mathbb{R}^n$. For every $\epsilon > 0$, there is a one-pass streaming algorithm that on an input stream vector $v \in \mathbb{R}^n$ computes, with probability at least 0.99, a $(1 \pm \epsilon)$-approximation to $l(v)$, and uses $\mathrm{mmc}(l)^2 \cdot \mathrm{poly}(\log(n)/\epsilon)$ bits of space.*

**Theorem 1.2** (Lower Bound)**.** *Let $l$ be a symmetric norm on $\mathbb{R}^n$. Any turnstile streaming algorithm that outputs, with probability at least 0.99, a $(1 \pm 1/6)$-approximation for $l(\cdot)$ must use $\Omega(\mathrm{mmc}(l)^2)$ bits of space in the worst case.*

For the coarser $D$-approximation, where $D > 1.1$ and can grow with $n$, in Theorem 1.3 we build upon the algorithm of Theorem 1.1 trading the larger approximation ratio for a $1/D^2$ multiplicative decrease in storage. It turns out that the quadratic dependence on $D$ is the best possible; we prove the matching lower bound in Theorem 1.4.

**Theorem 1.3.** *Let $l$ be a symmetric norm on $\mathbb{R}^n$. For every $1.1 \leq D \leq \mathrm{mmc}(l)$ there is a one-pass streaming algorithm that on input stream vector $v \in \mathbb{R}^n$ computes, with probability at least 0.99, a $D$-approximation to $l(v)$ and uses $(\mathrm{mmc}(l)^2/D^2) \cdot \mathrm{poly}(\log n)$ bits of space.*

**Theorem 1.4.** *Let $l$ be a symmetric norm on $\mathbb{R}^n$. Any turnstile streaming algorithm that outputs, with probability at least 0.99, a $D$-approximation for $l(\cdot)$ must use $\Omega(\mathrm{mmc}(l)^2/D^2)$ bits of space in the worst case.*

We prove the upper bound theorems in Sections 3 and 5, respectively, with some details in the full version of this paper.). The lower bounds both appear in Section 4. To our knowledge, this is the first application of measure concentration to streaming algorithms (Chernoff and Hoeffding bounds aside). The geometric and analytical properties of high-dimensional normed spaces have become well understood over decades of research. We hope that more tools from that field can be brought to bear on these intriguing streaming problems, see Section 7 for promising directions for further work.

**Applications and Examples.** Section 6 describes some applications of our results. One application is to a class of norms called $Q'$ norms [Bha97], which includes the $l_p$ norms for $1 \leq p \leq 2$, among others. $Q'$ norms are just the dual norms to $Q$ norms (shorthand for quadratic), which in turn are norms of the form $l(x) = \Phi(x^2)^{1/2}$, for some symmetric norm $\Phi$, where $x^2$ denotes coordinate-wise squaring of $x$. We study these norms in Section 6.2. The upshot is that every $Q'$ norm $l'$ has $\mathrm{mmc}(l') = O(\log n)$, and thus can be computed by a streaming algorithm using polylogarithmic space. Several $Q'$ norms have been proposed as regularizers for sparse recovery problems in Machine Learning. One such norm is the $k$-support norm [AFS12], which is more conveniently described via its unit ball $C_k = \mathrm{conv}\{x \in \mathbb{R}^n : |\mathrm{supp}(x)| \leq k \text{ and } l_2(x) \leq 1\}$. It is not readily apparent how to design a specialized streaming algorithm for this norm, but we obtain such an explicit algorithm, for every $k$, as a special case of $Q'$ norms. Another example is the box-$\Theta$ norm [MPS14], where given $0 < a < b \leq c$, we let $\Theta := \{\theta \in [a,b]^n : l_1(\theta) \leq c\}$, and define the box-$\Theta$ norm as

$$l_\Theta(x) := \min_{\theta \in \Theta} \Big(\sum_{i=1}^n x_i^2/\theta_i\Big)^{1/2}, \qquad \text{and its dual norm is} \qquad l'_\Theta(x) := \max_{\theta \in \Theta} \Big(\sum_{i=1}^n \theta_i x_i^2\Big)^{1/2}.$$

It's easy to see that every box-$\Theta$ norm is a $Q'$ norm, and therefore has polylogarithmic streaming space complexity. To the best of our knowledge, there is no other technique that can approximate these norms on a streaming vector.

Our results also apply to what we shall call the *top-$k$ norm*. Denoted as $\Phi_k(x)$, it is defined as the sum of the $k$ largest coordinates of $|x|$ [Bha97]. This norm is a special case of the Ky Fan $k$-norm and is sometimes studied as a toy example to understand regularization of the Ky Fan norms [WDST14]. We show in Section 6.1 that $\mathrm{mmc}(\Phi_k) = \tilde{\Theta}(\sqrt{n/k})$, so when $k$ is large, for example linear in $n$, the top-$k$ norm of a stream vector can be approximated in only polylogarithmic space. We are aware of no other streaming algorithms that can approximate this norm, as ours does.

## 1.2 Overview of Techniques

**Upper Bound.** Our algorithm for Theorem 1.1 uses a linear sketch in the style of Indyk and Woodruff's sketch for large frequency moments [IW05], but the size of the sketch is calibrated by $\mathrm{mmc}(l)$. The algorithm is presented in Section 3, with some details presented in the full version. This is a surprising application of the Indyk-Woodruff sketching technique, as all previous applications of this method are to computing functions with an additive structure $\sum_{i=1}^{n} f(v_i)$. In these settings, the Indyk-Woodruff algorithm can be viewed as performing Importance Sampling of the summands of the target function $\sum_{i=1}^{n} f(v_i)$. However, a symmetric norm $l$ need not have an explicit mathematical formula, let alone be decomposable as a sum, and we thus need a different way to identify the "important" coordinates, which informally means that zeroing these coordinates would introduce too much error to $l(v)$. At a high level, our analysis makes two major contributions. The first is to provide an explicit criterion for importance, and the second is to reveal that inside this importance criterion, the most crucial quantity is the maximum modulus of concentration $\mathrm{mmc}(l)$. A more detailed outline of the analysis follows, omitting constants and dependence on $\epsilon$.

First, we imagine rounding each coordinate of the streaming vector $v$ to a power of $\alpha = 1 + 1/\mathrm{polylog}(n)$, which can be seen to have negligible effect using basic properties of symmetric norms. Moreover, since the norm is symmetric, it suffices to know only the number of coordinates, $b_i$, at each "level" $\alpha^i$. By our choice of $\alpha$, there are only $\mathrm{polylog}(n)$ levels, so we can represent the rounded vector succinctly. Recovering the rounded vector exactly would require linear storage, so we use the Indyk-Woodruff sampling technique to approximate the vector.

The Indyk-Woodruff procedure approximates each $b_i$ by sampling each coordinate $i$ of the vector $v$ with probability $\mathrm{polylog}(n)/b_i$, and then in the sampled vector (which is expected to have $\mathrm{polylog}(n)$ coordinates of level $i$ whenever $b_i \neq 0$), the algorithm identifies $l_2$-heavy-hitters. If the coordinates of level $i$ are $l_2$-heavy-hitters in the sampled vector (they are in the same level and thus have about the same value), then we get a good estimate of $b_i$; it's not as simple as counting them and scaling inversely to the sampling probability, but that is the right idea. If the coordinates are not $l_2$-heavy-hitters, then we get no estimate for $b_i$, and must assume it is 0. We show that if we parameterize the sketch according to $\mathrm{mmc}(l)^2$, then we get approximations to all the "important" levels, which is sufficient to accurately recover $l(v)$.

**Lower Bound.** The lower bound of Theorem 1.2 is proved using a reduction from the Communication Complexity of multiparty set-disjointness, and concentration of measure of the norm $l$ again plays a key role. In the disjointness setting, each of $t$ players is given a subset of $[n]$, and their task is to determine whether the sets are mutually disjoint or are "uniquely" intersecting. Instead of the standard reduction, where each player places in the stream one update to $v_i$ for every element $i \in [n]$ in the set he holds, in our reduction, each player $j \in [t]$ adds to the stream a vector $w^{(i,j)} \in \mathbb{R}^n$ whenever element $i$ is in his set. Each vector $w^{(i,j)}$ is random but the entire collection of vectors is designed so that the resulting stream vector is, roughly, a uniformly random vector on a "disjoint" instance, and a vector maximizing the norm on an "intersecting" instance. For these two cases to be well-separated, we must choose the number of players $t$ to be large enough. By applying concentration of measure, we show that $t = O(\sqrt{n}/\mathrm{mmc}(l))$ players suffice, and, by known communication bounds for disjointness [CKS03, BJKS04, Gro09], this leads to an $\Omega(n/t^2) = \Omega(\mathrm{mmc}(l)^2)$ storage lower bound for every algorithm approximating the norm $l$ to within $1 \pm 1/6$ (the constant $1/6$ is arbitrary). Extending the lower bound to a $D$-approximation, for $D$ bounded away from 1, can be accomplished with the same reduction using $t = O(D\sqrt{n}/\mathrm{mmc}(l))$ players instead, which yields Theorem 1.4. The proofs of both lower bound theorems can be found in Section 4.

**Optimal Tradeoff.** For the $D$-approximation algorithm, Theorem 1.3, the idea is to define, given a norm $l$, a new symmetric function $l_{(D)} : \mathbb{R}^n \to \mathbb{R}_{\geq 0}$ such that $l(x) \leq l_{(D)}(x) \leq Dl(x)$. Even though $l_{(D)}$ is not a norm, we can still define $\mathrm{mmc}(l_{(D)})$, which is bounded as $\tilde{O}(\mathrm{mmc}(l)/D)$. The approximation comes by using our main algorithm to get a 1.1-approximation to $l_{(D)}(v)$, which translates into a $2D$-approximation of $l(v)$. The definition of $l_{(D)}$ and its analysis are presented in Section 5.

## 1.3 Related Work

There has been extensive work on computing norms, and related functions, in the sketching and streaming models. Most recently, Andoni, Krauthgamer, and Razenshteyn [AKR15] have shown that a normed space $(\mathbb{R}^n, l)$ embeds linearly into $l_{0.99}$ with distortion $D > 1$ if and only if this normed space admits distance estimation sketching with approximation $\Theta(D)$ and sketch size $O(1)$ bits. Thus, they characterize sketching of a general norm by its embeddability. In comparison, our characterization applies only to symmetric norms, but we consider streaming (not sketching) algorithms, which in Theorem 1.1 means a stronger consequence, and in Theorem 1.2 means a stronger assumption. And perhaps more importantly, our results achieve $(1+\epsilon)$-approximation, while their algorithm achieves approximation proportional to $D$ (though their lower bound shows a linear tradeoff with sketch size).

Another important tool that may seem relevant is that every turnstile streaming algorithm can be replaced by a linear sketch, as shown by Li, Nguyen, and Woodruff [LNW14]. However, this transformation does not make it easy to determine the streaming complexity of a given symmetric norm $l$, because it is not easy to design a linear sketch for $l$.

There are other generic streaming algorithms that provide approximation guarantees for an entire class of functions of the form $\sum_{i=1}^n f(v_i)$, where $f$ is some nonnegative function [BO10, BO13a]. If one has a so-called $f$-heavy-hitters algorithm that identifies every coordinate $i$ accounting for half of the total sum, i.e., $f(v_i) \geq \sum_{j \neq i} f(v_j)$ and moreover approximates this $f(v_i)$, then one can also approximate the sum $\sum_{i=1}^n f(v_i)$, incurring only an $O(\log n)$ factor overhead on top of the $f$-heavy-hitters algorithm's storage. For a large class of functions $f$, including monotone functions, computing $f$-heavy-hitters can be reduced to computing $l_2$-heavy-hitters in several random sub-streams [BOR15, BCWY16] or even just random sampling [BC15]. Universality falls out as a side-effect of the design of the algorithm — the only dependence on $f$ is through the number of sub-streams, which determines the sketch size, up to a polylog$(n)$ factor. Therefore, any two functions that lead to the same sketch size, in fact, use the exact same sketch.

Finally, we should mention there is a very long line of results on estimating $l_p$ norms (also called frequency moments) in a data stream, including designing small-space algorithms [AMS99, IW05, Ind06, GC07, Li08, KNW10, AKO11, BO13a, BO13b, BKSV14, Gan15] and proving space lower bounds [SS02, CKS03, BJKS04, Gro09, ANPW13, LW13]. This list omits improvements of the runtime of update and output procedures, and devising extensions like $l_p$ sampling.

## 2 Preliminaries

An important unit vector for us is $\xi^{(n')} := \frac{1}{\sqrt{n'}}(1, 1, 1, \ldots 1, 0, \ldots, 0) \in \mathbb{R}^n$, for any $n' \leq n$, which has $n'$ nonzero coordinates. We abuse the notation to write $\xi^{(n')} \in \mathbb{R}^{n'}$ by removing zero coordinates, and vice-versa by appending zeros. Let us record some basic facts about symmetric norms.

**Lemma 2.1** (Monotonicity of Symmetric Norms, see e.g. Proposition IV.1.1 in [Bha97]). *If $l(\cdot)$ is a symmetric norm and $x, y \in \mathbb{R}^n$ satisfy that for all $i$, $|x_i| \leq |y_i|$, then $l(x) \leq l(y)$.*

Without loss of generality, we assume that our norms are normalized on the standard basis, i.e., $l(e_i) = 1$. Recall that the *dual* of a norm $l : \mathbb{R}^n \to \mathbb{R}$ is the norm $l' : \mathbb{R}^n \to \mathbb{R}$ given by $l'(x) := \sup\{\frac{|\langle x, y \rangle|}{l(y)} : y \neq 0\}$. For the following facts see, e.g., [MS86, Sections 3.1.2 and 4.5].

**Fact 2.2.** *For all $x \in \mathbb{R}^n$, $l_\infty(x) \leq l(x) \leq l_1(x)$.*

**Fact 2.3.** *Let $a, b > 0$ be such that, for all $x \in \mathbb{R}^n$, $a^{-1} l_2(x) \leq l(x) \leq b\, l_2(x)$. Then, for all $x \in \mathbb{R}^n$, $b^{-1}\, l_2(x) \leq l'(x) \leq a\, l_2(x)$.*

**Fact 2.4.** $\mathrm{M}_l\, \mathrm{M}_{l'} \geq 1$.

We restrict attention to vectors $v$ whose coordinates are in the range $\{-m, \ldots, m\}$, for $m = \text{poly}(n)$, so $\log m = O(\log n)$. Our results still apply when $m$ is larger but one must replace $\log n$ factors with $\log m$ factors.

Last, we must be precise about the model of computation, because we do not have a mathematical formula for the norm. Our algorithm will rely on evaluating the norm on a vector that is derived from a sketch of the stream. Every coordinate of this vector should be easy to recover from the sketch, but the vector need not be written explicitly, to avoid $\Omega(n)$ storage. To accomodate this, we make the assumption that our algorithm has access to an oracle NORM that computes $l(v)$ using queries to the coordinates of $v$, i.e., our algorithm must provide query access to any coordinate $v_i$.

## 3 An Algorithm for Symmetric Norms

In this section we prove Theorem 1.1, which shows that a symmetric norm can be approximated in the turnstile streaming model using one pass and $O(\text{mmc}(l)^2 \text{poly}(1/\epsilon \cdot \log n))$ bits of memory. The Algorithm 1, uses a subroutine called `Level1`, whose full description appears in the full version. The rest of this section considers a given symmetric norm $l$ on $\mathbb{R}^n$ and a desired accuracy parameter $0 < \epsilon < 1$. Let the two parameters $\alpha > 1$ and $0 < \beta \leq 1$ be determined later, possibly depending on $n$, $\epsilon$ and $\text{mmc}(l)$. We assume $\text{mmc}(l) \leq \gamma \sqrt{n}$, for some sufficiently small constant $0 < \gamma \ll 1/2$, since otherwise the lower bound given in Theorem 1.2 implies that linear memory is necessary to approximate this norm with a streaming algorithm.

### 3.1 Level Vectors and Important Levels

**Definition 3.1** (Important Levels). *For $v \in \mathbb{R}^n$, define level $i$ as $B_i := \{j \in [n] : \alpha^{i-1} \leq |v_j| < \alpha^i\}$, and denote its size by $b_i := |B_i|$. We say that level $i$ is $\beta$-important if*

$$b_i > \beta \sum_{j>i} b_j; \quad and \quad b_i \alpha^{2i} \geq \beta \sum_{j \leq i} b_j \alpha^{2j}.$$

Recall from Section 2 that we restrict attention to vectors $v$ whose coordinates are in the range $\{-m, \ldots, m\}$, for $m = \text{poly}(n)$. This assumption implies that the number of non-zero $b_i$'s is at most $t = O(\log_\alpha n)$. And if we normalize $v$ to a unit vector in $l_2$-norm, then every non-zero coordinate has absolute value at least $1/\text{poly}(n)$.

We will rely on the next theorem, which shows a streaming algorithm recovers all the important $b_i$'s. Its proof appears in the full version.

**Theorem 3.2.** *For every $\epsilon > 0$, there is a one-pass streaming algorithm `Level1` that given an input stream and parameters $\alpha' = 1 + \gamma > 1$ and $0 < \beta \leq 1$, outputs $\{\hat{b}_i\}$ for base $\alpha = 1 + O(\gamma)$, such that with probability $1 - O(1/\text{poly}(n))$, for all $i$,*

- $\hat{b}_i \leq b_i$; and
- *if level $i$ is $\beta$-important, then $\hat{b}_i \geq (1-\epsilon) b_i$.*

*This algorithm uses $O(\gamma^{-5} \epsilon^{-2} \beta^{-1} \log^{12} n)$ bits of space.*

To state and analyze our algorithm for approximating $l(v)$, we introduce the following notation. Later, we shall omit $(v)$ from the notation, as it is clear from the context.

**Definition 3.3** (Level Vectors and Buckets). *Define the level vector for $v \in \mathbb{R}^n$ with integer coordinates to be*

$$V(v) := (\underbrace{\alpha^1, \ldots, \alpha^1}_{b_1 \text{ times}}, \underbrace{\alpha^2, \ldots, \alpha^2}_{b_2 \text{ times}}, \ldots, \underbrace{\alpha^t, \ldots, \alpha^t}_{b_t \text{ times}}, 0, \ldots, 0) \in \mathbb{R}^n;$$

*and define the $i$-th bucket of $V(v)$ to be*

$$V_i(v) := (\underbrace{0, \ldots, 0}_{b_1+b_2+\ldots+b_{i-1} \text{ times}}, \underbrace{\alpha^i, \ldots, \alpha^i}_{b_i \text{ times}}, \underbrace{0, \ldots, 0}_{b_{i+1}+b_{i+2}\ldots b_t \text{ times}}, 0, \ldots, 0) \in \mathbb{R}^n.$$

*Let $\hat{V}(v)$ and $\hat{V}_i(v)$ be defined similarly for the approximated values $\{\hat{b}_i\}$. We denote $V(v) \setminus V_i(v)$ as the vector with the $i$-th bucket replaced by $0$; and denote $V(v) \setminus V_i(v) \cup \hat{V}_i(v)$ as the vector by replacing the whole $i$-th bucket with $\hat{V}_i(v)$, i.e.,*

$$V(v) \setminus V_i(v) \cup \hat{V}_i(v) := (\underbrace{\alpha^1, \ldots, \alpha^1}_{b_1 \text{ times}}, \ldots, \underbrace{\alpha^i, \ldots, \alpha^i}_{\hat{b}_i \text{ times}}, \ldots, \underbrace{\alpha^t, \ldots, \alpha^t}_{b_t \text{ times}}, 0, \ldots, 0) \in \mathbb{R}^n.$$

## 3.2 Approximated Levels Provide a Good Approximation

We first show the level vector $V$ can be used to approximate $l(v)$, if we choose a base $\alpha := 1 + O(\epsilon)$.

**Proposition 3.4.** *For all $v \in \mathbb{R}^n$, $l(V(v))/\alpha \leq l(v) \leq l(V(v))$.*

*Proof.* Follows directly from the monotonicity of symmetric norms (Lemma 2.1). □

The next key lemma shows that $l(\hat{V})$ is a good approximation to $l(V)$.

**Lemma 3.5** (Bucket Approximation). *For every level $i$, if $\hat{b}_i \leq b_i$, then $l(V \setminus V_i \cup \hat{V}_i) \leq l(V)$; and if $\hat{b}_i \geq (1-\epsilon)b_i$, then $l(V \setminus V_i \cup \hat{V}_i) \geq (1-\epsilon)l(V)$.*

*Proof.* The upper bound follows immediately from the monotonicity of norms. We will prove the lower bound as follows. Let us take the vector
$$\hat{V}_i := (\underbrace{0,0,\ldots 0}_{b_1+b_2+\ldots+b_{i-1}\text{ times}}, \underbrace{\alpha^i,\ldots \alpha^i}_{\hat{b}_i\text{ times}}, 0,\ldots,0).$$

Let us also define $W := V - V_i$. Note that $W + \hat{V}_i$ is a permutation of a vector $V \setminus V_i \cup \hat{V}_i$. We will prove that, under assumptions of the lemma, $l(W + \hat{V}_i) \geq (\hat{b}_i/b_i)l(V)$.

For a vector $v \in \mathbb{R}^n$ and a permutation $\pi \in \Sigma_n$, we denote $\pi(v)$ a vector in $\mathbb{R}^n$ such that $\pi(v)_i := v_{\pi(i)}$. Since the norm $l$ is symmetric, we have that $l(v) = l(\pi(v))$. Consider a set of permutations $S$, consisting of all permutations that are cyclic shifts over the non-zero coordinates of $V_i$, and do not move any other coordinates. That is, there is exactly $b_i$ permultations in $S$, and for every $\pi \in S$, we have $\pi(W) = W$. By the construction of the set $S$, we have,
$$\sum_{\pi \in S} \pi(\hat{V}_i) = \hat{b}_i V_i$$
and therefore $\sum_{\pi \in S} \pi(W + \hat{V}_i) = \hat{b}_i V_i + b_i W$. As vectors $V_i$ and $W$ have disjoint support, by monotonicity of the norm $l$ with respect to each coordinates we can deduce $l(\hat{b}_i V_i + b_i W) \geq l(\hat{b}_i(V_i + W))$. By plugging those together,
$$\hat{b}_i l(V_i + W) \leq l(\hat{b}_i V_i + b_i W) = l\left(\sum_{\pi \in S} \pi(\hat{V}_i + W)\right)$$
$$\leq \sum_{\pi \in S} l\left(\pi(\hat{V}_i + W)\right) = b_i l(\hat{V}_i + W) \tag{1}$$

where the last equality follows from the fact that $l$ is symmetric and $|S| = b_i$. Hence, $l(\hat{V}_i + W) \geq \frac{\hat{b}_i}{b_i} l(V) \geq (1-\epsilon)l(V)$, as desired. □

## 3.3 Contributing Levels and Important Levels

**Definition 3.6** (Contributing Levels). *Level $i$ is called $\beta$-contributing if $l(V_i) \geq \beta\, l(V)$.*

**Lemma 3.7.** *Let $V'$ be the vector obtained from $V$ by removing all levels that are not $\beta$-contributing. Then $(1 - O(\log_\alpha n) \cdot \beta)l(V) \leq l(V') \leq l(V)$.*

*Proof.* Let $i_1,\ldots,i_k$ be the levels that are not $\beta$-contributing. Then by the triangle inequality,
$$l(V) \geq l(V) - l(V_{i_1}) - \ldots - l(V_{i_k}) \geq (1 - k\beta)l(V).$$
The proof follows by bounding $k$ by $t = O(\log_\alpha n)$, which is the total number of non-zero $b_i$'s. □

The following lemma and Lemma 3.15 show together that that every $\beta$-contributing level is also $\beta'$-important for a suitable $\beta'$ that depends on $\mathrm{mmc}(l)$.

**Lemma 3.8.** *If level $i$ is $\beta$-contributing, then $b_i \geq \frac{\lambda \beta^2}{\mathrm{mmc}(l)^2 \log^2 n} \cdot \sum_{j>i} b_j$ for some absolute constant $\lambda > 0$.*

We present the following concentration of measure results for the proof of this lemma,

**Lemma 3.9.** *For every norm $l$ on $\mathbb{R}^n$, if $x \in S^{n-1}$ is drawn uniformly at random according to Haar measure on the sphere, then*
$$\Pr(|l(x) - \mathrm{M}_l| > \frac{2\,\mathfrak{b}_l}{\sqrt{n}}) < \frac{1}{3}$$

**Lemma 3.10.** *For every $n > 0$, there is a vector $x \in S^{n-1}$ satisfying*

1. $|l_\infty(x) - M_{l_\infty^{(n)}}| \leq 2/\sqrt{n}$,
2. $|l(x) - M_{l^{(n)}}| \leq 2\,\mathfrak{b}_{l^{(n)}}/\sqrt{n}$, and
3. $|\{i : |x_i| > \frac{1}{K\sqrt{n}}\}| > \frac{n}{2}$ *for some universal constant $K$.*

We prove these lemmas using Levy's isoperimetric inequality, see e.g. [MS86, Section 2.3].

**Theorem 3.11** (Levy's Isoperimetric Inequality)**.** *For a continuous function $f : S^{n-1} \to \mathbb{R}$, let $M_f$ be the median of $f$, i.e., $\mu(\{x : f(x) \leq M_f\}) \geq 1/2$ and $\mu(\{x : f(x) \geq M_f\}) \geq 1/2$, where $\mu(\cdot)$ is the Haar probability measure on the unit sphere $S^{n-1}$. Then $\mu(\{x : f(x) = M_f\}_\epsilon) \geq 1 - \sqrt{\pi/2}\,e^{-\epsilon^2 n/2}$, where for a set $A \subset S^{n-1}$ we denote $A_\epsilon := \{x : l_2(x, A) \leq \epsilon\}$ and $l_2(x, A) := \inf_{y \in A} \|x - y\|_2$.*

*Proof of Lemma 3.9.* By applying Theorem 3.11, for random $x$ distributed according to the Haar measure on the $l_2$-sphere, with probability at least $1 - \sqrt{\pi/2}\,e^{-2} > \frac{2}{3}$ there is some $y \in S^{n-1}$, such that $\|x - y\|_2 \leq \frac{2}{\sqrt{n}}$ and $l(y) = M_l$. We know that norm $l$ is $\mathfrak{b}_l$-Lipschitz with respect to $\|\cdot\|_2$, and as such

$$|l(x) - M_l| = |l(x) - l(y)| \leq l(x - y) \leq \mathfrak{b}_l \|x - y\| \leq \frac{2\,\mathfrak{b}_l}{\sqrt{n}}$$
$\square$

*Proof of Lemma 3.10.* Consider $x$ drawn uniformly at random from a unit sphere. According to Lemma 3.9, we have $\Pr(|l_\infty(x) - M_{l_\infty^{(n)}}| > 2/\sqrt{n}) < \frac{1}{3}$ and $\Pr(|l(x) - M_l| > 2\mathfrak{b}_l/\sqrt{n}) < \frac{1}{3}$.

Let us define $\tau(x, t) := |\{i : |x_i| < t\}|$. We need to show that for some universal constant $K$, with probability larger than $\frac{2}{3}$ over a choice of $x$, we have $\tau(x, \frac{1}{K\sqrt{n}}) < \frac{n}{2}$.

Indeed, consider random vector $z \in \mathbb{R}^n$, such that all coordinates $z_i$ are independent standard normal random variables. It is well known, that $\frac{z}{\|z\|_2}$ is distributed uniformly over a sphere, and therefore has the same distribution as $x$. There is a universal constant $K_1$ such that $\Pr(\|z\|_2 > K_1\sqrt{n}) < \frac{1}{6}$, and similarly, there is a constant $K_2$, such that $\Pr(|z_i| < \frac{1}{K_2}) < \frac{1}{12}$. Therefore, by Markov bound we have $\Pr(\tau(z, \frac{1}{K_2}) > \frac{n}{2}) < \frac{1}{6}$. Using union bound, with probability larger than $\frac{2}{3}$ it holds simultaneously that $\|z\|_2 \leq K_1\sqrt{n}$ and $\tau(z, \frac{1}{K_2}) < \frac{n}{2}$, in which case $\tau(z/\|z\|_2, \frac{1}{K_1 K_2 \sqrt{n}}) < \frac{n}{2}$.

Finally, by union bound, a random vector $x$ satisfies all of the conditions in the statement of the lemma with positive probability. $\square$

We now prove that the norm $l$ of the (normalized) all-ones vector $\xi^{(n)}$ is closely related to the median of the norm. This all-ones vector is useful because it can be easily related to a single level of $V$.

**Lemma 3.12** (Flat Median Lemma)**.** *Let $l : \mathbb{R}^n \to \mathbb{R}$ be a symmetric norm. Then*

$$\lambda_1 M_l / \sqrt{\log n} \leq l(\xi^{(n)}) \leq \lambda_2 M_l,$$

*where $\lambda_1, \lambda_2 > 0$ are absolute constants.*

Note that the first inequality is tight for $l_\infty$. To prove this lemma, we will need the following well-known fact, see e.g. [MS86].

**Fact 3.13.** *There are absolute constants $0 < \gamma_1 \leq \gamma_2$ such that for every integer $n \geq 1$,*
$$\gamma_1 \sqrt{\log(n)/n} \leq M_{l_\infty^{(n)}} \leq \gamma_2 \sqrt{\log(n)/n}.$$

*Proof of Lemma 3.12.* Using Lemma 3.10, there is a constant $\lambda > 0$ and a vector $x \in S^{n-1}$ such that (i) $|l_\infty(x) - M_{l_\infty}| \leq \lambda \sqrt{1/n}$, (ii) $|l(x) - M_l| \leq \lambda \mathfrak{b}_l / \sqrt{n}$ and (iii) $|\{i : |x_i| > \frac{1}{K\sqrt{n}}\}| > \frac{n}{2}$. By Fact 3.13, $M_{l_\infty} = \Theta(\sqrt{\log(n)/n})$. On the other hand, $\mathrm{mmc}(l) \leq \gamma \sqrt{n}$, for sufficiently small $\gamma$, thus $\lambda \mathfrak{b}_l / \sqrt{n} < M_l$. We can therefore get constants $\gamma_1, \gamma_2 > 0$ such that $\gamma_1 M_l \leq l(x) \leq \gamma_2 M_l$ and $\gamma_1 \sqrt{\log(n)/n} \leq l_\infty(x) \leq \gamma_2 \sqrt{\log(n)/n}$. Therefore $|x| \leq \gamma_2 \sqrt{\log n}\,\xi^{(n)}$ coordinate-wise, and by monotonicity of symmetric norms,

$$\gamma_1 M_l \leq l(x) \leq \gamma_2 \sqrt{\log n}\, l(\xi^{(n)}). \qquad (2)$$

For the second part of the lemma, let $J = \{i : |x_i| > \frac{1}{K\sqrt{n}}\}$. As $|J| > \frac{n}{2}$, there is a permutation $\pi$ such that $[n] - J \subset \pi(J)$. Let $|x|$ be a vector obtained from $x$ by taking an absolute value of every coordinate, and let $\pi(x)$ denote applying permutation $\pi$ to coordinates of vector $x$. We have $|x| + \pi(|x|) > \frac{\xi^{(n)}}{K}$ coordinate-wise, and therefore by monotonicity of symmetric norms, we have

$$\frac{1}{K} l(\xi^{(n)}) \leq l(|x| + \pi(|x|)) \leq l(|x|) + l(\pi(|x|)) = 2l(x) \leq 2\gamma_2 M_l.$$
$\square$

Next, we show that the median is roughly monotone (in $n$), which is crucial for the norm to be approximated.

**Lemma 3.14** (Monotonicity of Median)**.** *Let $l : \mathbb{R}^n \to \mathbb{R}$ be a symmetric norm. For all $n' \leq n'' \leq n$,*
$$\mathrm{M}_{l^{(n')}} \leq \lambda \operatorname{mmc}(l)\sqrt{\log n'}\, \mathrm{M}_{l^{(n'')}},$$
*where $\lambda > 0$ is an absolute constant.*

*Proof.* By Lemma 3.12 and the fact that $\xi^{(n')}$ is also a vector in $S^{n''-1}$,
$$\lambda \mathrm{M}_{l^{(n')}} / \sqrt{\log n'} \leq l(\xi^{(n')}) \leq \mathfrak{b}_{l^{(n'')}} \leq \operatorname{mmc}(l) \mathrm{M}_{l^{(n'')}}.$$
□

We are now ready to prove the Lemma 3.8.

*Proof of Lemma 3.8.* Fix a $\beta$-contributing level $i$, and let $U$ be the vector $V$ after removing buckets $j = 0, \ldots, i$. By Lemma 3.12, there is an absolute constant $\lambda_1 > 0$ such that
$$l(V_i) = \alpha^i \sqrt{b_i}\, l(\xi^{(b_i)}) \leq \lambda_1 \alpha^i \sqrt{b_i}\, \mathrm{M}_{l^{(b_i)}},$$
and similarly
$$l(U) \geq \frac{\lambda_2 \alpha^i}{\sqrt{\log n}} \sqrt{\sum_{j>i} b_j}\, \mathrm{M}_{l^{(\sum_{j>i} b_j)}}.$$

We now relate these two inequalities as follows. First, $l(V_i) \geq \beta\, l(V) \geq \beta\, l(U)$. Second, we may assume $b_i < \sum_{j>i} b_j$, as otherwise the lemma holds, and then by monotonicity of the median (Lemma 3.14) $\mathrm{M}_{l^{(b_i)}} \leq \lambda_3 \operatorname{mmc}(l)\sqrt{\log n}\, \mathrm{M}_{l^{\sum_{j>i} b_j}}$, for some absolute constant $\lambda_3 > 0$. Putting these together, we get
$$\beta \cdot \frac{\lambda_2 \alpha^i}{\sqrt{\log n}} \sqrt{\sum_{j>i} b_j} \leq \lambda_1 \alpha^i \sqrt{b_i} \cdot \lambda_3 \operatorname{mmc}(l)\sqrt{\log n},$$
and the lemma follows. □

**Lemma 3.15.** *If level $i$ is $\beta$-contributing, then there is an absolute constant $\lambda > 0$ such that*
$$b_i \alpha^{2i} \geq \frac{\lambda \beta^2}{\operatorname{mmc}(l)^2 (\log_\alpha n) \log^2 n} \sum_{j \leq i} b_j \alpha^{2j}.$$

*Proof of Lemma 3.15.* Fix a $\beta$-contributing level $i$, and let $h := \operatorname{argmax}_{j \leq i} \sqrt{b_j} \alpha^j$. We proceed by separating into two cases. First, if $b_i \geq b_h$ then the lemma follows easily by
$$\sum_{j \leq i} b_j \alpha^{2j} \leq t b_h \alpha^{2h} \leq O(\log_\alpha n) b_i \alpha^{2i}.$$
The second case is when $b_i < b_h$. Using the definition of a contributing level and Lemma 3.12,
$$\lambda_1 \alpha^i \sqrt{b_i}\, \mathrm{M}_{l^{(b_i)}} \geq l(V_i) \geq \beta l(V) \geq \lambda_2 \beta \alpha^h \sqrt{b_h / \log n}\, \mathrm{M}_{l^{(b_h)}},$$
for some absolute constants $\lambda_1, \lambda_2 > 0$. Plugging in $\mathrm{M}_{l^{(b_i)}} \leq \lambda_3 \operatorname{mmc}(l)\sqrt{\log n}\, \mathrm{M}_{l^{(b_h)}}$, which follows from monotonicity of the median (Lemma 3.14), for some absolute constant $\lambda_3 > 0$, we get
$$\lambda_1 \alpha^i \sqrt{b_i}\, \mathrm{M}_{l^{(b_i)}} \geq \frac{\lambda_2 \beta \sqrt{b_h} \alpha^h}{\sqrt{\log n}} \cdot \frac{\mathrm{M}_{l^{(b_i)}}}{\lambda_3 \operatorname{mmc}(l)\sqrt{\log n}},$$
$$\sqrt{b_i} \alpha^i \geq \frac{\lambda_2 \beta \sqrt{b_h} \alpha^h}{\lambda_1 \lambda_3 \operatorname{mmc}(l) \log n}.$$
Squaring the above and observing that $b_h \alpha^{2h} \geq \frac{1}{O(\log_\alpha n)} \sum_{j \leq i} b_j \alpha^{2j}$, the proof is complete. □

## 3.4 Putting It Together

*Proof of Theorem 1.1.* Recall from Section 2 that we assume our algorithm has access to an oracle `NORM` that computes $l(v)$ using queries to the coordinates of $v$, i.e., our algorithm must provide query access to any coordinate $v_i$. We assume without loss of generality that $\epsilon \geq 1/\operatorname{poly}(n)$, because an exact algorithm using space $O(n \log n)$ is trivial.

Our algorithm maintains a data structure that eventually produces a vector $\hat{V}$. We will show that with high probability, $l(\hat{V})$ approximates $l(v)$ and we will also bound the space required for the data structure. The algorithm is presented in Algorithm 1. The idea is to run the `Level1` algorithm with appropriate parameters. Specifically, to achieve $(1 \pm \epsilon)$-approximation to $l(v)$, we set the approximation guarantee of the buckets to be $\epsilon' := O\left(\frac{\epsilon^2}{\log n}\right)$ and the importance guarantee to be $\beta' := O\left(\frac{\epsilon^5}{\mathrm{mmc}(l)^2 \log^5 n}\right)$.

---

**Algorithm 1** `OnePassSymmetricNorm`$(\mathcal{S}, n)$

1: **Input:** stream $\mathcal{S}$ of from domain $[n]$, and $\epsilon > 0$
2: **Output:** $X$
3: $(\alpha, \hat{b}_1, \hat{b}_2, \ldots, \hat{b}_t) \leftarrow \texttt{Level1}(\mathcal{S}, n, \alpha' = 1 + O(\epsilon), \epsilon' = O\left(\frac{\epsilon^2}{\log n}\right), \beta' = O\left(\frac{\epsilon^5}{\mathrm{mmc}(l)^2 \log^5 n}\right), \delta = \frac{0.01\epsilon}{n})$;
4: Construct $\hat{V}$ using $\alpha$ and $\hat{b}_1, \hat{b}_2, \ldots, \hat{b}_t$;
5: Invoke `NORM`, answer each query for $v_i$ by $\hat{V}_i$;
6: $X \leftarrow$ output of `NORM`.
7: Return $X$.

---

Let $v$ be the streaming vector. It is approximated by its level vector $V$ with base $\alpha = 1 + O(\epsilon)$, namely, $(1 - O(\epsilon))l(v) \leq l(V) \leq l(v)$ by Proposition 3.4. Observe that $t = O(\log_\alpha n) = O(\log(n)/\epsilon)$, and assume that algorithm `Level1` succeeds, i.e., the high-probability event in Theorem 3.2 indeed occured. Denote by $\hat{V}$ the output of `Level1`, and by $V'$ the vector $V$ after removing all buckets that are not $\beta$-contributing, and define $\hat{V}'$ similarly to $\hat{V}$, where we set $\beta := \epsilon/t = O(\epsilon^2/\log n)$. Every $\beta$-contributing level is necessarily $\beta'$-important by Lemmas 3.8 and 3.15 and therefore satisfies $\hat{b}_i \geq (1 - \epsilon')b_i$. We bound the error from removing non-contributing levels by Lemma 3.7, namely,
$$(1 - O(\epsilon))\, l(V) \leq (1 - O(\log_\alpha n) \cdot \beta)\, l(V) \leq l(V') \leq l(V).$$
By monotonicity (Lemma 2.1) and by Lemma 3.5,
$$l(\hat{V}) \geq l(\hat{V}') = l((V' \setminus V_{i_1} \cup \hat{V}_{i_1}) \ldots \setminus V_{i_k} \cup \hat{V}_{i_k})$$
$$\geq (1 - \epsilon')^t\, l(V') \geq (1 - O(\epsilon))l(V').$$
Altogether, $(1 - O(\epsilon))l(v) \leq l(\hat{V}') \leq l(v)$, which bounds the error of $l(\hat{V}')$ as required.

The space requirement of the algorithm is dominated by that of subroutine `Level1`, namely, $O\left(\frac{\log^{12} n}{\beta' \epsilon'^2 \epsilon^5}\right) = O\left(\frac{\mathrm{mmc}(l)^2 \log^{19} n}{\epsilon^{14}}\right)$ bits. Storing the data structure, i.e., $\hat{b}_i$'s, requires only
$O(\log_\alpha n) \log n = O\left(\frac{\log^2 n}{\epsilon}\right)$ bits. $\square$

## 4 Lower Bound

The overall plan is to use the multiparty disjointness communication complexity problem to prove an $\Omega(\mathrm{mmc}(l)^2)$ bits storage lower bound on any turnstile streaming algorithm outputs a $(1 \pm 1/6)$-approximation, or better, to the norm of the frequency vector. The bound is otherwise independent of the norm or $n$.

Multiparty disjointness is a communication problem where there are $t$ players who each recieve a subset of $[n]$, and their goal is to determine whether their sets are intersecting or not. The problem was introduced by Alon, Matias, and Szegedy [AMS99] to prove storage lower bounds for the frequency moments problem. After several improvements [CKS03, BJKS04], the communication complexity of multiparty disjointness was settled at an asymptotically optimal $\Omega(n/t)$ bits of communication by Gronemeier [Gro09].

### 4.1 John's Theorem for Symmetric Norms

We will start by proving the following specialization of John's Theorem [Joh48] to the case of symmetric norms.

**Theorem 4.1** (John's Theorem for Symmetric Norms). *If $l(\cdot)$ is a symmetric norm on $\mathbb{R}^n$, then there exist $0 < a \leq b$ such that $b/a \leq \sqrt{n}$ and, for all $x \in \mathbb{R}^n$, $al_2(x) \leq l(x) \leq bl_2(x)$.*

*Proof.* By John's Theorem [Joh48] there exists a unique ellipsoid $E$ of maximum volume contained in $B = \{x \in \mathbb{R}^n \mid l(x) \leq 1\}$ and, furthermore, $B \subseteq \sqrt{n}E$. $E$ is permutation and sign symmetric because $B$ is, so it follows from Lemma 4.2 that $E$ is a sphere. Therefore, there exist $0 < a < b$ such that $al_2(x) \leq l(x) \leq bl_2(x)$, for all $x \in \mathbb{R}^n$, and, furthermore, $b/a \leq \sqrt{n}$. $\square$

**Lemma 4.2.** *If an ellipsoid $E$ is symmetric under every permutation or change of signs to its coordinates then $E$ is a sphere.*

*Proof.* Let $A$ be a positive semidefinite matrix such that $E = \{x \in \mathbb{R}^n \mid x^T A x = 1\}$. Since $A$ is a real positive semidefinite matrix, it can be decomposed as $A = SDS^T$, where $S$ is orthonormal and $D$ is a diagonal matrix with $D_{11} \geq D_{22} \geq \cdots D_{nn} \geq 0$. We will show that all of the diagonal entries in $D$ are the same, from which it follows that $A = D$ and $E$ is a sphere. Let $s_i$, for $i \in [n]$, be the columns of $S$. Let $i \neq 1$, choose a permutation $P_1$ so that $P_1 s_1$ has its coordinates in decreasing order by magnitude, and choose a permutation $P_i$ so that $P_i s_i$ has the same. Now choose a diagonal matrix $D$ that has $D_{jj} = 1$ if $(P_1 s_1)_j$ has the same sign as $(P_i s_i)_j$, and $D_{jj} = -1$ if the signs are different, zeros may be treated arbitrarily. Let $P = P_1^T D P_i$; since $P$ is the product of permutation matrics and a sign change matrix we have $E = \{x \mid x^T P A P^T x \leq 1\}$ by the symmetry assumption.

We have $D_{11} = s_1^T A s_1$, since $s_1$ is a unit vector orthogonal to $s_i$, $i > 1$. Let $\lambda = S^T P^T s_1$. By construction we have $\sum_j \lambda_j^2 = 1$ and $\lambda_i > 0$. If we suppose that $D_{ii} < D_{11}$, then we arrive at the following contradiction

$$D_{11} = s_1 A s_1 = s_1 P A P^T s_1 = \sum_{j=1}^n \lambda_j^2 D_{jj} < D_{11}.$$

Therefore, $D_{11} = D_{ii}$, for all $i$, and $E$ is sphere. $\square$

## 4.2 Concentration of a Symmetric Norm

Let us begin by discussing a concentration inequality for symmetric norms. We will need concentration of $l(Z)$ around $\sqrt{n}\, \mathrm{M}_l$, where $Z$ is distributed according to the canonical Gaussian disribution on $n$ dimensions. To get it, we will use the following two concentration theorems for Lipschitz functions. The diffrence between them is the underlying distribution, whether it is uniform on $S^{n-1}$ or multivariate Gaussian. Comparing $l(Z)$ against its own median is just a direct application of Theorem 4.4. There is a little bit more work to do because we wish to compare $l(Z)$ to the median of $l(\cdot)$ over $S^{n-1}$, which is also the median of $l(Z)/l_2(Z)$. Note that the $M$ in Theorem 4.3 is not the same as the $M$ in Theorem 4.4 because the probability distributions are different.

**Theorem 4.3** ([MS86]). *Let $f : S^{n-1} \to \mathbb{R}$ be 1-Lipschitz, let $Z \in S^{n-1}$ be chosen uniformly at random, and let $M$ be the median of $f(Z)$. Then, for all $t > 0$, $\Pr(|f(Z) - M| \geq t) \leq 2 e^{-nt^2/2}$.*

**Theorem 4.4** ([LT13]). *Let $f : \mathbb{R}^n \to \mathbb{R}$ be 1-Lipschitz, let $Z_1, Z_2, \ldots, Z_n \overset{\text{iid}}{\sim} N(0,1)$, and let $M$ be the median of $f(Z)$. Then, for all $t > 0$, $\Pr(|f(Z) - M| \geq t) \leq e^{-t^2/2}$.*

It will also be helpful to have the following fact about $\chi^2$ random variables.

**Lemma 4.5.** *([LM00]) Let $X \sim \chi_n^2$. For all $x \geq 0$,*
$$\Pr(X \geq n + 2\sqrt{nx} + x) \leq e^{-x} \quad \text{and} \quad \Pr(X \leq n - 2\sqrt{nx}) \leq e^{-x}.$$

**Lemma 4.6.** *Let $n \geq 2$ and let $Z \in \mathbb{R}^n$ be a random vector with coordinates $Z_1, Z_2, \ldots, Z_n \overset{\text{iid}}{\sim} N(0,1)$. Let $\mathrm{M}_l$ be the median of $l(\cdot)$ on $S^{n-1}$, where $l(\cdot)$ is a symmetric norm on $\mathbb{R}^n$. Then, for all $t \geq 0$,*
$$\Pr(|l(Z) - \sqrt{n}\, \mathrm{M}_l\,| \geq t\sqrt{n}\, \mathrm{M}_l) \leq 7 e^{-t^2/200}.$$

*Proof.* We first establish an inequality that does not have the correct dependence on $t$, it is (4), and then use it to bound the median of $l(Z)$ in terms of $\sqrt{n}\, \mathrm{M}_l$. That will allow us to apply Theorem 4.4 and get the bound above.

By Theorem 4.1, there exist $0 < a_l \leq b_l$ such that $b_l/a_l \leq \sqrt{n}$ and, for all $x \in \mathbb{R}^n$, $a_l l_2(x) \leq l(x) \leq b_l l_2(x)$. This implies $l(\cdot)$ is $b_l$-Lipschitz on $\mathbb{R}^n$. By scaling the norm (and, as a consequence, $\mathrm{M}_l$), we may assume $a_l = 1$ without loss of generality.

It is easy to see that
$$\Pr(l(Z) - \sqrt{n}\, \mathrm{M}_l \geq t\sqrt{n}\, \mathrm{M}_l)$$
$$= \Pr\left(l(Z) - l_2(Z)\, \mathrm{M}_l + l_2(Z)\, \mathrm{M}_l - \sqrt{n}\, \mathrm{M}_l \geq t\sqrt{n}\, \mathrm{M}_l\right)$$
$$\leq \Pr\left(l(Z) - l_2(Z)\, \mathrm{M}_l \geq \sqrt{n}\, \mathrm{M}_l\, \frac{t}{2}\right) + \Pr\left(l_2(Z) - \sqrt{n} \geq \sqrt{n}\, \frac{t}{2}\right). \tag{3}$$

For the second term, notice that $l_2(Z)^2$ is a $\chi_n^2$ random variable. Using Lemma 4.5, we have
$$\Pr\left(l_2(Z) - \sqrt{n} \geq \sqrt{n}\frac{t}{2}\right) = \Pr\left(l_2(Z)^2 \geq n(1+\frac{t}{2})^2\right)$$
$$= \Pr\left(l_2(Z)^2 \geq n + 2\sqrt{n}(\sqrt{n}\frac{t}{2}) + (\sqrt{n}\frac{t}{2})^2\right) \leq e^{-nt^2/4}.$$

For the first term in (3), we have
$$\Pr\left(|l(Z) - l_2(Z)\operatorname{M}_l| \geq \sqrt{n}\operatorname{M}_l\frac{t}{2}\right)$$
$$\leq \Pr\left(\left|l(\frac{Z}{l_2(Z)}) - \operatorname{M}_l\right| \geq \operatorname{M}_l\frac{t}{4}\right) + \Pr(l_2(Z) \geq 2\sqrt{n}).$$

The scaled norm $l(\cdot)/\mathfrak{b}_l$ is 1-Lipschitz and $Z/l_2(Z)$ is distributed according to the Haar distribution, so by Theorem 4.3 and our previous $\chi^2$ bound we have
$$\Pr\left(l(Z) - l_2(Z)\operatorname{M}_l \geq \sqrt{n}\operatorname{M}_l\frac{t}{2}\right) \leq 2\exp\{-\frac{n\operatorname{M}_l^2 t^2}{32\,\mathfrak{b}_l^2}\} + e^{-n}$$
$$\leq 2\exp\{-\frac{t^2}{32}\} + e^{-n}$$
where the final inequality follows because $\operatorname{M}_l / \mathfrak{b}_l \geq a_l/b_l \geq 1/\sqrt{n}$.

So far, we have established, $\forall\, t \geq 0$,
$$\Pr(l(Z) - \sqrt{n}\operatorname{M}_l \geq t\sqrt{n}\operatorname{M}_l) \leq 2e^{-t^2/32} + e^{-n} + e^{-nt^2/4}. \tag{4}$$
It is almost the bound that we want, except for the $e^{-n}$ term. Substituting in $t = 8$ and $n \geq 2$ we find $\Pr(l(Z) \geq 9\sqrt{n}\operatorname{M}_l) \leq \frac{1}{2}$. Therefore the median of $l(Z)$ is at no larger than $9\sqrt{n}\operatorname{M}_l$, so Theorem 4.4 implies, $\forall\, t \geq 0$ and $n \geq 2$,
$$\Pr(l(Z) - 9\sqrt{n}\operatorname{M}_l \geq t\sqrt{n}\operatorname{M}_l) \leq e^{-t^2 n \operatorname{M}_l^2/2\,\mathfrak{b}_l^2} \leq e^{-t^2/2}. \tag{5}$$
The last step is to combine these two bounds by using (4) to bound,
$$\forall t \leq 10 \text{ and } n \geq 2,\ \Pr(l(Z) \geq t\sqrt{n}\operatorname{M}_l) \leq 3e^{-t^2/32} + e^{-n} \leq 7e^{-t^2/32}$$
and using (5) to establish, $\forall t \geq 10$ and $n \geq 2$,
$$\Pr(l(Z) \geq t\sqrt{n}\operatorname{M}_l) = \Pr(l(Z) - 9\sqrt{n}\operatorname{M}_l \geq (t-9)\sqrt{n}\operatorname{M}_l)$$
$$\leq \Pr(l(Z) - 9\sqrt{n}\operatorname{M}_l \geq \frac{t}{10}\sqrt{n}\operatorname{M}_l) \leq e^{-t^2/200},$$
which proves the theorem. □

### 4.3 The Norm of a Randomized Vector

The multiparty disjointness reduction used to prove Theorem 1.2 uses a randomized vector. Given a vector $v \in \mathbb{R}^n$, we randomize it by replacing the coordinates by independent Normally distributed random variables $V_i \sim N(0, v_i^2)$, for each $i \in [n]$.

The next lemma allows us to compare the distribution of the norm of two different randomized vectors. Recall that a random variable $Y$ is said to *stochastically dominate* a random variable $X$ if $\Pr(Y \geq t) \geq \Pr(X \geq t)$ for all $t \in \mathbb{R}$, or, equivalently, their cdf's satisfy $F_X \geq F_Y$.

**Lemma 4.7.** *Let $\sigma, \tau \in \mathbb{R}_{\geq 0}^n$ satisfy $\sigma \leq \tau$ coordinate-wise. Let $X_i \sim N(0, \sigma_i^2)$, independently for $i = 1, \ldots, n$, and $Y_i \sim N(0, \tau_i^2)$, independently for $i = 1, \ldots, n$. Then $l(Y)$ stochastically dominates $l(X)$, in particular, for all $t \in \mathbb{R}$,*
$$\Pr(l(X) \geq t) \leq \Pr(l(Y) \geq t).$$

*Proof.* It is well known that, for any random variables $Y'$ and $X'$, $Y'$ stochastically dominates $X'$ if and only if there is a coupling of $X'$ and $Y'$ so that $X' \leq Y'$. Since $\tau_i \geq \sigma_i$ we have that $|Y_i|$ stochastically dominates $|X_i|$, for all $i$. Therefore, there is a coupling of the vectors $|X|$ and $|Y|$ so that $|X| \leq |Y|$ coordinate-wise at every sample point. This is also a coupling of $l(X)$ and $l(Y)$, and by applying Lemma 2.1 proves that $l(X) \leq l(Y)$ at every sample point. Thus, $l(Y)$ stochastically dominates $l(X)$. □

The main technical lemma we use to prove Theorem 1.2 is the following.

**Lemma 4.8.** *If $v \in S^{n-1}$ has $l(v) = \mathfrak{b}_l$ and $V \in \mathbb{R}^n$ is a random vector with coordinates distributed $V_i \stackrel{\text{iid}}{\sim} N(0, v_i^2)$, then $\Pr(l(V) \geq \mathfrak{b}_l/4) \geq 1/10$.*

In order to prove Lemma 4.8 we will first need to bound $\mathbb{E}\, l(V)$.

**Lemma 4.9.** *If $v \in S^{n-1}$ has $l(v) = \mathfrak{b}_l$ and $V$ is a random vector with coordinates distributed $V_i \stackrel{\text{iid}}{\sim} N(0, v_i^2)$, then $\mathbb{E}\, l(V) \geq 0.49\, \mathfrak{b}_l$.*

*Proof.* If on every outcome it happened that $|V| \geq |v|$ coordinate-wise then Lemma 2.1 would imply the desired result. Of course, it is very likely that for some coordinates $|V_i| < |v_i|$. The idea of the proof is to "patch up" those coordinates with another vector that has small norm and then apply the reverse triangle inequality. Let $U = \max\{|v| - |V|, 0\}$, where the maximum is taken coordinate-wise. $U$ was chosen so that $|V| + U \geq |v|$, hence by Lemma 2.1 $l(|V| + U) \geq l(v) = \mathfrak{b}_l$, and by the reverse triangle inequality $l(V) \geq l(v) - l(U)$.

It remains to bound $\mathbb{E}\, l(U)$. We will begin by bounding $\mathbb{E}\, l_2(U)$ and use this value to bound $\mathbb{E}\, l(U)$. Let $Z \sim N(0,1)$ and let $\mathbf{1}_A$ be the indicator function of the set $A$. Direct calculation with the Normal c.d.f. shows that

$$\mathbb{E}\, l_2(U)^2 = \sum_{i=1}^n 2v_i^2\, \mathbb{E}\left(\mathbf{1}_{[0,1)}(Z)\,(1-Z)^2\right) \leq 0.26 \sum v_i^2 = 0.26,$$

Therefore, $\mathbb{E}\, l_2(U) \leq \left(\mathbb{E}\, l_2(U)^2\right)^{1/2} \leq 0.51$, where the first inequality is Jensen's. Finally, we can conclude $\mathbb{E}\, l(U) \leq \mathfrak{b}_l\, \mathbb{E}\, l_2(U) \leq 0.51\, \mathfrak{b}_l$ and $\mathbb{E}\, l(V) \geq l(v) - \mathbb{E}\, l(U) \geq 0.49\, \mathfrak{b}_l$. □

*Proof of Lemma 4.8.* For a random variable $X$ and event $A$, let $\mathbb{E}(X; A) = \mathbb{E}\, X\mathbf{1}_A(X) = \int_A X\, dP$. We begin with the trivial bound, for any $0 < \alpha < \beta$,

$$\mathbb{E}\, l(V) = \mathbb{E}(l(V); (0, \alpha]) + \mathbb{E}(l(V); (\alpha, \beta]) + \mathbb{E}(l(V); (\beta, \infty))$$
$$\leq \alpha + \beta \Pr(l(V) \in (\alpha, \beta]) + \mathbb{E}(l(V); (\beta, \infty)). \tag{6}$$

We shall use $l_2(V)$ to bound the last term above. Observe that $\mathbb{E}\, l_2(V)^2 = 1$ and, letting $Z_1, \ldots, Z_n \stackrel{\text{iid}}{\sim} N(0,1)$,

$$\text{Var}(l_2(V)^2) = \text{Var}\left(\sum_i v_i^2 Z_i^2\right)$$
$$= \sum_i v_i^4 \text{Var}(Z_i^2) = 2\sum_i v_i^4 \leq 2l_2(v)^2 = 2,$$

because $v \in S^{n-1}$ has unit length. For $k > 0$, we have by Chebyshev's Inequality that

$$\Pr(l_2(V)^2 \geq \sqrt{2}k + 1) \leq \frac{1}{k^2},$$

and, by a change of variables,

$$\Pr(l_2(V) \geq x) \leq \left(\frac{x^2 - 1}{\sqrt{2}}\right)^{-2} = \frac{2}{(x^2-1)^2} \leq 4/x^4, \quad \text{for } x > \sqrt{2},$$

and it extends trivially to all $x > 0$. This implies $\Pr(l(V) \geq \mathfrak{b}_l x) \leq 4/x^4$, hence $\Pr(l(V) \geq x) \leq 4(\mathfrak{b}_l/x)^4$. Thus,

$$\mathbb{E}(l(V); (\beta, \infty)) \leq \int_\beta^\infty \frac{4\,\mathfrak{b}_l^4}{x^4} dx = \frac{4\,\mathfrak{b}_l^4}{3\beta^3}.$$

Now we return to (6) and substitute $\alpha = 0.49\, \mathfrak{b}_l/4$ and $\beta = 2.44\, \mathfrak{b}_l$. Together with Lemma 4.9 we get

$$0.49\, \mathfrak{b}_l \leq \mathbb{E}\, l(V) \leq \frac{0.49\, \mathfrak{b}_l}{4} + 2.32\, \mathfrak{b}_l \Pr(l(V) \geq \mathfrak{b}_l/4) + \frac{4\,\mathfrak{b}_l}{3(2.44)^3}.$$

Upon rearranging the inequality we find $\Pr(l(V) \geq \mathfrak{b}_l/4) \geq 1/10$, as desired. □

## 4.4 Multiparty Disjointness and the Norm on a Stream

We will show an $\Omega(\text{mc}(l)^2)$ bits bound on the storage needed by a streaming algorithm for the norm $l$.

**Lemma 4.10.** *Let $l(\cdot)$ be a symmetric norm on $\mathbb{R}^n$. A turnstile streaming algorithm that outputs a $(1 \pm \frac{1}{6})$-approximation for $l(\cdot)$, with probability at least 0.99, uses $\Omega(\text{mc}(l)^2)$ bits in the worst case.*

Let recall that every symmetric norm $l(\cdot)$ on $\mathbb{R}^n$ induces the norm $l(\cdot)^{(k)}$ on $\mathbb{R}^k$, for $k < n$, by setting any $n - k$ coordinates to 0. The induced norm may have a different ratio of $\mathfrak{b}_l / \mathrm{M}_l$. Since a streaming algorithm that approximates $l(\cdot)$ must also approximate $l(\cdot)^{(k)}$, Lemma 4.10 in fact implies Theorem 1.2.

*Proof of Lemma 4.10.* We begin with an instance of the multiparty disjointness promise problem on domain $[n]$ with $t = \lceil 240\sqrt{n} \cdot \mathrm{M}_l / \mathfrak{b}_l \rceil$ players. By Lemma 4.1, $t \geq 240\sqrt{n}\,\mathrm{M}_l / \mathfrak{b}_l \geq 240$. The players are given sets $P_1, P_2, \ldots, P_t \subseteq [n]$ with the promise that either they are pairwise disjoint or exactly one element is contained in every set but they are otherwise disjoint. The players are allowed, in any order, to communicate bits with each other by writing them to a shared blackboard, and they are given shared access to a string of random bits. The players' goal is for at least one among them to determine whether the sets $P_1, \ldots, P_t$ are disjoint or intersecting. If the players correctly determine the type of instance with probability at least 0.55, then their communication scheme is called a "correct protocol". It is known that for any correct protocol, the players must write $\Omega(n/t)$ bits to the blackboard in the worst case [CKS03]. In this reduction, each of the $t$ players will transmit the memory of the streaming algorithm once, which leads to an $\Omega(n/t^2) = \Omega(\mathfrak{b}_l^2 / \mathrm{M}_l^2)$ bits lower bound on the memory used by the algorithm.

Next, we describe the protocol under the assumption that the players can perform computations with real numbers. After describing the protocol we explain that this assumption can be removed by rounding the real values to a sufficiently high precision.

The players have shared access $n^2$ i.i.d. $N(0,1)$ random variables $Z_{i,j}$, for $i, j \in [n]$, and additional independent randomness for the approximation algorithm.

Let $v \in \mathrm{argmax}_{x \in S^{n-1}} l(x)$, so that $l(v) = \mathfrak{b}_l$. We define an $n \times n$ matrix $V$ with coordinates
$$V_{i,j} = Z_{i,j} v_{i+j \bmod n}.$$
Since $v$ is fixed, all of the players can compute the matrix using the shared randomness. Let $V_j$ denote the $j$th column of $V$; it is a vector with independent Normally distributed entries. The vector of standard deviations of $V_j$ is a copy of $v$ that has been cyclically shifted down by $j$ entries, in particular the standard deviation of $V_{i,n}$ is $v_i$.

Here is the stream that the players create, they jointly run an approximation algorithm for the norm on this stream. For each player $i$ and item $j \in P_i$ the $i$th player adds a copy of $V_j$ to the stream. More precisely, for each $j \in P_i$ player $i$ adds 1 with frequency $V_{1,j}$, 2 with frequency $V_{2,j}$, etc. The players repeat this protocol 10 times independently.

Now we analyze the possible outcomes of one of the ten trials. Let $N \subseteq \cup_{i=1}^t P_i$ be the set of elements that appear in exactly one set $P_i$, and let $X = \sum_{j \in N} V_j$. If there is no intersection between the $P_i$'s, then the stream's frequency vector is $X$. If they all intersect at $j^*$, then the frequency vector is $Y = tV_{j^*} + X$.

It remains to compare $l(X)$ and $l(Y)$. The coordinates of $X$ are independent and normally distributed with zero mean and variance
$$\mathbb{E}\,X_i^2 = \sum_{j \in N} \mathbb{E}\,V_{i,j}^2 = \sum_{j \in N} v_{i+j \bmod n}^2 \leq \sum_{j=1}^n v_j^2 = 1.$$
Let $Z$ be a random vector with coordinates $Z_i \overset{\mathrm{iid}}{\sim} N(0,1)$, for $i = 1, \ldots, n$. By Lemma 4.7 $Z$ stochastically dominates $X$, and using Lemma 4.6 we have $\Pr(\frac{1}{\sqrt{n}} l(X) \geq 40\,\mathrm{M}_l) \leq \Pr(\frac{1}{\sqrt{n}} l(Z) \geq 40\,\mathrm{M}_l) \leq 0.005$. On the other hand, $Y$ stochastically dominates $tV_{j^*}$ and Lemma 4.8 additionally implies
$$\Pr(l(tV_{j^*} + X) \geq 60\,\mathrm{M}_l\,\sqrt{n}) \geq \Pr(tl(V_{j^*}) \geq 60\,\mathrm{M}_l\,\sqrt{n}) \geq \Pr(l(V_{j^*}) \geq \mathfrak{b}_l\,/4) \geq 1/10.$$
The final player checks whether the maximum approximation returned among the 10 trials is larger or smaller than $50\,\mathrm{M}_l\,\sqrt{n}$ and declares "intersecting" or "disjoint" accordingly.

The output of the protocol is correct on an intersecting instance if at least one of the 10 stream vectors has norm larger than $60\,\mathrm{M}_l\,\sqrt{n}$ and the algorithm always returns a $(1 \pm 1/6)$-approximation. It is correct on a disjoint instance if all of the stream vectors have norm smaller than $40\,\mathrm{M}_l\,\sqrt{n}$ and the algorithm always returns a $(1 \pm 1/6)$-approximation. If the instance is an intersecting one, then with probability at least 0.1 the magnitude of $l(tV_{j^*} + X)$ is large enough. At least one of the ten trials will have this property with probability at least $1 - 0.9^{10} \geq 0.65$, because the trials use independent random matrices. Since the algorithm correctly approximates the norm with probability at least 0.99, it follows that the protocol is correct for an intersecting instance with probability at least $0.65 - 10 \cdot 0.01 = 0.55$.

On a disjoint instance, one trial of the protocol is successful with probability at least $0.99^2 \geq 0.98$ where one factor comes from the success of the approximation algorithm and the other from our earlier application of the concentration bound. Thus, the output of the protocol correctly identifies a disjoint instance with probability at least $1 - 10 \cdot 0.02 = 0.8$, by a union bound over the ten trials. Therefore, this protocol is a correct protocol.

It remains to describe the rounding of the real values. It suffices to represent each value with a sufficiently high precision. We replace each variable as $Z_{i,j}$ with a discrete random variable $Z_{i,j} = \hat{Z}_{i,j} + \delta_{i,j}$ where $\hat{Z}_{i,j}$ are distributed i.i.d. $N(0,1)$ and $\delta_{i,j}$ is difference between $\hat{Z}_{i,j}$ and its closest point in $\{\frac{j}{n^4} \mid j = -n^5, \ldots, n^5 - 1, n^5\}$. In particular, with very high probability, $|\delta_{i,j}| \leq 1/2n^4$ for all pairs $i, j$. We also replace $v$ by a vector $v = \hat{v} + \delta_v$ where $\hat{v} \in \mathrm{argmax}_{x \in S^{n-1}} l(x)$, so that $l(\hat{v}) = \mathfrak{b}_l$, and where $\delta_v$ is a vector containing the difference between each entry of $\hat{v}_i$ and the nearest integer multiple of $n^{-4}$ to it.

Each frequency in the stream is the sum of at most $t$ variables. Performing these replacements changes each frequency in the stream by no more than $2tn^{-4}$. Let $\Delta \in \mathbb{R}^n$ denote this change, then $l(\Delta) \leq \mathfrak{b}_l \, l_2(\Delta) \leq 2\mathfrak{b}_l \, tn^{-7/2} = O(\mathrm{M}_l/n^3)$. Applying the triangle and reverse triangle inequalities shows that the change negligible. Therefore, the discretized protocol is correct also, which completes the proof. $\square$

Suppose there is an algorithm with the weaker, $D$-approximation guarantee. Namely, $D > 1$ and with probability at least 0.99, the algorithm returns a value $\hat{l}$ satisfying $l(V) \leq \hat{l} \leq Dl(V)$, where $V$ is the stream vector. The main lower bound, Theorem 1.2, can be easily adapted this setting, where we get a lower bound of $\Omega(\mathrm{mmc}(l)^2/D^2)$ bits instead, with a small modification to the proof of Lemma 4.10.

**Theorem 1.4.** *Let $l$ be a symmetric norm on $\mathbb{R}^n$. Any turnstile streaming algorithm that outputs, with probability at least 0.99, a $D$-approximation for $l(\cdot)$ must use $\Omega(\mathrm{mmc}(l)^2/D^2)$ bits of space in the worst case.*

Indeed, the proof goes as above, except that the number of players should be increased to $t = \lceil 240 D \sqrt{n} \, \mathrm{M}_l / \mathfrak{b}_l \rceil$. The disjoint instances do not change, but the norm is $D$ times larger on an intersecting instance. Thus, the $D$-approximation algorithm can distinguish the two and we get the bound $\Omega(\mathrm{mc}(l)^2/D^2)$, which is easily boosted to $\Omega(\mathrm{mmc}(l)^2/D^2)$ bits, as before. When $l = l_\infty$, this matches the trade-off proved by Saks and Sun [SS02].

## 5 Optimal Space-Approximation Tradeoff

In this section we obtain a nearly tight space-approximation tradeoff for computing any symmetric norm in the data-stream model. Specifically, we show below how our earlier streaming algorithm can be adapted to match the lower bound in Theorem 1.4, up to a polylog($n$) factor. The adapted algorithm achieves, for any $D \geq 1.1$ and symmetric norm $l$, a $D$-approximation within $\tilde{O}(\mathrm{mmc}(l)^2/D^2)$ bits of storage. The key part of the analysis is to define, a new symmetric function $l_{(D)}$ on $\mathbb{R}^n$ such that $l(x) \leq l_{(D)}(x) \leq Dl(x)$, for all $x \in \mathbb{R}^n$, and such that our earlier algorithm can find a $(1 \pm 1/2)$-approximation to $l'(x)$ using polylog($n$) $\cdot \mathrm{mmc}(l)^2/D^2$ bits of space.

We start in Section 5.1 with an algorithm for $Q$-norms (formaly defined in Section 6.2), a special case that is easier to prove. We then leverage ideas from this simpler case to design in Section 5.2 an algorithm for general symmetric norm.

### 5.1 $D$-Approximation for $Q$-norms

**Theorem 5.1.** *Let $l : \mathbb{R}^N \to \mathbb{R}$ be a $Q$-norm. Then for every $1.1 < D \leq \mathrm{mmc}(l)$ there is a randomized streaming algorithm that $D$-approximates $l$ and uses $\tilde{O}(\mathrm{mmc}(l)^2/D^2)$ bits of space.*

*Proof.* Fix a $Q$-norm $l$ and $1 < D \leq \mathrm{mmc}(l)$. We first show that for all $x \in \mathbb{R}^N$,
$$l_{(D)}(x) := \max\left(\frac{D \, \mathrm{M}_l \, l_2(x)}{\log n}, l(x)\right)$$
is an $O(D)$-approximation to $l(x)$. Since $l$ is a $Q$-norm, we have by Lemma 6.9 that $\xi^{(n)}$ is roughly a minimizer of $l(x)$ over $S^{N-1}$, namely,
$$\forall x \in \mathbb{R}^{N-1}, \qquad l(\xi^{(n)}) \, l_2(x) \leq 6\sqrt{\log n} \, l(x).$$
Recalling from Lemma 3.12 that, for some absolute constants $\lambda_1, \lambda_2 > 0$, $\lambda_1 \, \mathrm{M}_l/\sqrt{\log n} \leq l(\xi^{(n)}) \leq \lambda_2 \, \mathrm{M}_l$, we have that $\lambda_1 \, \mathrm{M}_l \, l_2(x) \leq 6(\log n) \cdot l(x)$. Altogether we obtain (assuming, without loss of generality, that $\lambda_1 < 1$)
$$\forall x \in \mathbb{R}^{N-1}, \qquad l(x) \leq l_{(D)}(x) \leq \frac{6D}{\lambda_1} l(x). \tag{7}$$

Our algorithm for $l$ simply applies Theorem 1.1 to compute an $O(1)$-approximation to $l_{(D)}(x)$, using $\mathrm{mmc}(l_{(D)})^2 \cdot \mathrm{polylog}(n)$ bits of space. This is indeed possible because $l_{(D)}$ is clearly a symmetric norm on $\mathbb{R}^n$, and yields an $O(D)$-approximation for $l$, which implies $D$-approximation by scaling $D$ appropriately.

It remains to bound $\mathrm{mmc}(l_{(D)})$ and show it is smaller than $\mathrm{mmc}(l)$ by factor $D$ roughly. By Lemma 6.8, there is an absolute constant $\lambda > 0$ such that $\mathrm{M}_{l^{(n')}} \geq \mathrm{M}_{l^{(n)}}/(\lambda\sqrt{\log n})$ for all $n' \leq n$. Let $n^* \leq n$ be such that $\mathrm{mmc}(l) = \mathfrak{b}_{l^{(n^*)}}/\mathrm{M}_{l^{(n^*)}}$, thus $\mathrm{mmc}(l) \leq \lambda\sqrt{\log n}\,\mathrm{mc}(l^{(n)})$. Since $D \leq \mathrm{mmc}(l)$, we have
$$D\,\mathrm{M}_l \leq \lambda\sqrt{\log n}\,\mathrm{mc}(l^{(n)})\,\mathrm{M}_l \leq \lambda\sqrt{\log n}\cdot\mathfrak{b}_l \quad \Rightarrow \quad \mathfrak{b}_{l_D^{(n')}} \leq \max\left(\lambda\mathfrak{b}_l/\sqrt{\log n},\mathfrak{b}_l\right).$$
By definition of $l_D$, $\mathrm{M}_{l_D^{(n')}} \geq \max\left(\frac{D\,\mathrm{M}_l}{\log n}, \frac{\mathrm{M}_l}{\lambda\sqrt{\log n}}\right)$. Thus,
$$\mathrm{mmc}(l_{(D)}) \leq \frac{\lambda \log n}{D}\mathrm{mc}(l^{(n)}) \leq \frac{\lambda \log n}{D}\mathrm{mmc}(l^{(n)}).$$

We conclude that there exists a streaming algorithm computes an $O(D)$-approximation for $l$ using $\tilde{O}(\mathrm{mmc}(l_D)^2) = \tilde{O}(\mathrm{mmc}(l)^2/D^2)$ bits of space. $\square$

## 5.2 $D$-Approximation for General Symmetric Norms

**Theorem 1.3.** *Let $l$ be a symmetric norm on $\mathbb{R}^n$. For every $1.1 \leq D \leq \mathrm{mmc}(l)$ there is a one-pass streaming algorithm that on input stream vector $v \in \mathbb{R}^n$ computes, with probability at least $0.99$, a $D$-approximation to $l(v)$ and uses $(\mathrm{mmc}(l)^2/D^2) \cdot \mathrm{poly}(\log n)$ bits of space.*

*Proof.* Let $\alpha > 1$ be a constant. Given a vector $v \in \mathbb{R}^n$ with integer coordinates, analogously to Defintion 3.3, define $V^\alpha = V_1^\alpha + V_2^\alpha + \ldots + V_t^\alpha$, where $V_i^\alpha$ is the level $i$ vector of $v$ with base $\alpha$, and appropriate $t = O(\log n)$. For each $i \in [t]$, we define similarly $b_i^{(\alpha)}$ as the number of coordinates falling into level $i$. Define for each integer $1 \leq n' \leq n$, $h(\xi^{(n')}) := \min\{Dl(\xi^{(n')}), \mathfrak{b}_{l^{(n')}}\}$, and
$$h(V_i^\alpha) := h(\xi^{(b_i^{(\alpha)})})l_2(V_i^\alpha) = \min\{Dl(V_i^\alpha), \mathfrak{b}_{l^{(b)}}l_2(V_i^\alpha)\}, \qquad \text{and}$$
$$h^{(\alpha)}(v) := \sum_{i \in [t]} h(V_i^\alpha).$$
We will omit the superscript $\alpha$ if it is clear from the context. We first claim that $h(v)$ is an $\tilde{O}(D)$-approximation to $l(v)$. Indeed
$$l(v) \leq \alpha \sum_{i \in [t]} l(V_i) \leq \alpha \sum_{i \in [t]} \min\{Dl(V_i), \mathfrak{b}_{l^{(b_i)}} l_2(V_i)\} = \alpha h(v), \tag{8}$$
and by monotonicity and homogeneity of norm $l$
$$h(v) = \sum_{i \in [t]} h(V_i) \leq \sum_{i \in [t]} Dl(V_i) \leq Dt \max_{i \in [t]} l(V_i) \leq (\lambda D \log n) l(v),$$
where $\lambda > 0$ is a constant. Thus $h(v)$ is an $\tilde{O}(D)$-approximation to $l(v)$.

It remains to prove that $h(v)$ can be $O(1)$-approximated using $\tilde{O}(\mathrm{mmc}(l)^2/D^2)$ bits of space. Let $\beta = O(1/\log n)$ and
$$\beta' = O\left(\frac{D^2 \beta^2}{\log^2 n\,\mathrm{mmc}(l)^2}\right).$$

Let $v \in \mathbb{R}^n$ be the streaming vector. We run algorithm `Level1` with importance parameter $\beta'$, base parameter $\alpha$ and constant error parameter $\epsilon \in (0, 1/2)$. By Theorem 3.2, `Level1` is guaranteed to output a vector $\hat{V}^{\alpha'}$ with base $\alpha' = \Theta(1)$ and with the following guarantees. Let $t' = O(\log n/\log \alpha')$, then for every $i \in [t']$, $\hat{b}_i^{(\alpha')} \leq b_i^{(\alpha')}$ and if $\hat{V}_i^{\alpha'}$ is $\beta'$-important, then also $(1-\epsilon)b_i^{(\alpha')} \leq \hat{b}_i^{(\alpha')}$. Thus,
$$\sum_{i \in [t']} h(\hat{V}_i^{\alpha'}) = \sum_{i \in [t']} \min\{Dl(\hat{V}_i^{\alpha'}), \mathfrak{b}_{l^{(\hat{b}_i)}} l_2(\hat{V}_i^{\alpha'})\} \leq \sum_{i \in [t']} \min\{Dl(V_i^{\alpha'}), \mathfrak{b}_{l^{(b_i)}} l_2(V_i^{\alpha'})\} = h(V^{\alpha'}).$$
We prove in Lemma 5.3 below that a $\beta$-contributing level of $h(v)$ (defined as $h(V_i^{\alpha'}) \geq \beta h(V^{\alpha'})$) is a $\beta'$-important level. Let $U \subset [t']$ be the set of contributing levels. Then,
$$h(\hat{V}^{\alpha'}) \geq \sum_{i \in U} h(\hat{V}_i^{\alpha'}) = \sum_{i \in U} \min\{Dl(\hat{V}_i^{\alpha'}), \mathfrak{b}_{l^{(\hat{b}_i)}} l_2(\hat{V}_i^{\alpha'})\} \geq \frac{(1-\epsilon)}{2} \sum_{i \in U} \min\{Dl(V_i^{\alpha'}), \mathfrak{b}_{l^{(b_i)}} l_2(V_i^{\alpha'})\},$$
where the second inequality follows from Lemma 3.5 and that $\mathfrak{b}_{l_{\hat{b}_i}} \geq \mathfrak{b}_{l_{b_i}}/2$. Indeed, let $v^* \in \mathbb{R}^{b_i}$, then we cut $v^*$ into two pieces with roughly equal number of non-zeros $v^* = v_1^* + v_2^*$, then $l(v^*) \leq l(v_1^*) + l(v_2^*) \leq 2\mathfrak{b}_{l_{\hat{b}_i}}$. On the other hand, $\sum_{i \notin U} h(V_i) \leq t\beta h(V) \leq \lambda_1 h(V)$, for some constant $\lambda_1 > 0$ that can be chosen arbitrarily small. Thus $h(\hat{V}^{\alpha'}) \geq (1-\epsilon)(1-\lambda_1)h^{(\alpha')}(v)/2$ is a constant-factor approximation to $h(V)$. Last, by Theorem 3.2, Level1 uses $\tilde{O}(1/\beta') = \tilde{O}(\mathrm{mmc}(l)^2/D^2)$ bits of space. □

**Lemma 5.2.** *For every integers $0 < n_1 \leq n_2 \leq n$,*
$$h(\xi^{(n_1)}) \leq \frac{\lambda\sqrt{\log n}\,\mathrm{mmc}(l)}{D} h(\xi^{(n_2)}),$$
*for some absoute constant $\lambda > 0$.*

*Proof.* Since $h(\xi^{(n_1)}) = \min\left(Dl(\xi^{(n_1)}), \mathfrak{b}_{l^{(n_1)}}\right)$ and $h(\xi^{(n_2)}) = \min\left(Dl(\xi^{(n_2)}), \mathfrak{b}_{l^{(n_2)}}\right)$, then
$$h(\xi^{(n_1)})/h(\xi^{(n_2)})$$
$$= \max\left(\frac{\min\left(Dl(\xi^{(n_1)}), \mathfrak{b}_{l^{(n_1)}}\right)}{Dl(\xi^{(n_2)})}, \frac{\min\left(Dl(\xi^{(n_1)}), \mathfrak{b}_{l^{(n_1)}}\right)}{\mathfrak{b}_{l^{(n_2)}}}\right)$$
$$\leq \max\left(\lambda\sqrt{\log n}\min\left(\mathrm{mmc}(l), \frac{\mathrm{mmc}(l)}{D}\right), \min\left(\frac{\lambda D \mathrm{M}_{l^{(n_1)}}}{\mathfrak{b}_{l^{(n_2)}}}, \frac{\mathfrak{b}_{l^{(n_1)}}}{\mathfrak{b}_{l^{(n_2)}}}\right)\right)$$
$$\leq \frac{\lambda\sqrt{\log n}\,\mathrm{mmc}(l)}{D}, \tag{9}$$
where the second inequality follows from Lemma 3.12 and Lemma 3.14. The last inequality uses $\mathfrak{b}_{l^{(n_1)}} \leq \mathfrak{b}_{l^{(n_2)}}$ and $\lambda > 0$ is an absolute constant. □

**Lemma 5.3.** *If a level $i$ is $\beta$-contributing, i.e., $h(V_i) \geq \beta h(V)$, then*

1. $b_i \geq \frac{\lambda D^2 \beta^2}{\log^2 n\, \mathrm{mmc}(l)^2} \sum_{j > i} b_j$;

2. $b_i \alpha^{2i} \geq \frac{\lambda D^2 \beta^2}{\log^2 n\, \mathrm{mmc}(l)^2} \sum_{j \leq i} b_j \alpha^{2j}$,

*for some constant $\lambda > 0$.*

*Proof.* The proof is similar to that of Lemma 3.8 and that of Lemma 3.15. Since level $i$ is $\beta$-contributing, we have
$$h(V_i) \geq \beta \sum_{j \in [t]} h(V_j).$$

Let $j^* = \mathrm{argmax}_{j > i} b_j$. We can assume $b_i \leq b_{j^*}$ since otherwise $b_i \geq \sum_{j > i} b_i/t$. Thus, by Lemma 5.2
$$h(V_i) \geq \beta h(V_{j^*}) \Rightarrow \sqrt{b_i} \geq \frac{D\beta}{\sqrt{\lambda' \log n}\,\mathrm{mmc}(l)} \sqrt{b_{j^*}} \Rightarrow b_i \geq \frac{D^2\beta^2}{\lambda' t \log n\,\mathrm{mmc}(l)^2} \sum_{j > i} b_j.$$
where $\lambda' > 0$ is an absolute constant.

For the second inequality, let $j' := \mathrm{argmax}_{j \leq i} \sqrt{b_j}\alpha^j$. We proceed by separating into two cases. First, if $b_i \geq b_{j'}$ then the lemma follows easily by
$$b_i \alpha^{2i} \geq b_{j'}\alpha^{2j'} \geq \frac{\sum_{j \leq i} b_j \alpha^{2j}}{t}.$$
The second case is when $b_i < b_{j'}$,
$$\alpha^i \sqrt{b_i} h(\xi^{(i)}) = h(V_i) \geq \beta h(V) \geq \beta h(V_{j'}) = \beta\alpha^{j'}\sqrt{b_{j'}} h(\xi^{(j')}).$$
By Lemma 5.2, we get
$$\alpha^i \sqrt{b_i} \geq \beta\alpha^{j'}\sqrt{b_{j'}} \frac{h(\xi^{(j')})}{h(\xi^{i})} \geq \frac{D\beta\sqrt{b_{j'}}\alpha^{j'}}{\sqrt{\lambda'' \log n}\,\mathrm{mmc}(l)},$$
where $\lambda'' > 0$ is an absolute constant. Squaring the above and observing that
$$b_i \alpha^{2i} \geq \frac{D^2\beta^2}{\lambda'' t \log n\,\mathrm{mmc}(l)^2} \sum_{j \leq i} b_j \alpha^{2j},$$
the proof is complete. □

# 6 Applications & Examples

## 6.1 The Top-$k$ Norm $\Phi_{(k)}$

The *top-$k$ norm* on $\mathbb{R}^n$ is simply the sum of the $k$ largest coordinates in absolute value, formally, $\Phi_{(k)}(x) := \sum_{i=1}^{k} |x|_{[i]}$, where $|x|_{[1]} \geq \ldots \geq |x|_{[n]}$ are the coordinates ordered by non-increasing absolute value. It is known (see e.g. [Bha97, Exer. IV.1.18]) that the dual norm of $\Phi_{(k)}$ is $\Phi'_{(k)}(x) := \max\{l_\infty(x), l_1(x)/k\}$. We can understand the streaming space complexity of such a norm $l$ by comparing the maximum and the median of such a norm over $S^{n-1}$, which is an easy calculation, and then applying Theorems 1.1 and 1.2.

**Theorem 6.1.** *There are absolute constants $\lambda_1, \lambda_2 > 0$ such that for all $k = 1, \ldots, n$,*

$$\lambda_1 \sqrt{\frac{n}{k \log n}} \leq \mathrm{mmc}(\Phi_{(k)}) \leq \lambda_2 \sqrt{\frac{n}{k}}, \qquad \text{and } \lambda_1 \sqrt{\frac{k}{\log(k)+1}} \leq \mathrm{mmc}(\Phi'_{(k)}) \leq \lambda_2 \sqrt{k}.$$

The above inequalities are existentially tight, by considering the cases $k = 1$ and $k = n$. To prove this theorem, we will need the next two lemmas. They both assume $1 \leq k \leq n$.

**Lemma 6.2.** *For all $x \in \mathbb{R}^n$, $\sqrt{\frac{k}{n}}\, l_2(x) \leq \Phi_{(k)}(x) \leq \sqrt{k}\, l_2(x)$.*

We remark that the second inequality above is tight for $x = \xi^{(k)}$.

*Proof.* Fix $x \in \mathbb{R}^n$. We use Cauchy-Schwarz

$$\sum_{i=1}^{k} |x|_{[i]} \leq \sqrt{k} \left( \sum_{i=1}^{k} |x|_{[i]}^2 \right)^{1/2} \leq \sqrt{k} \left( \sum_{i=1}^{n} x_{[i]}^2 \right)^{1/2}.$$

For the second inequality, we use monotonicity of $l_p$-norms

$$\sum_{i=1}^{k} |x|_{[i]} \geq \left( \sum_{i=1}^{k} |x|_{[i]}^2 \right)^{1/2} \geq \left( \frac{k}{n} \sum_{i=1}^{n} |x|_{[i]}^2 \right)^{1/2}. \qquad \square$$

**Lemma 6.3.** $\frac{\lambda_1 k}{\sqrt{n}} \leq \mathrm{M}_{\Phi_{(k)}} \leq \frac{\lambda_2 k \sqrt{\log n}}{\sqrt{n}}$ *for some absolute constants $\lambda_1, \lambda_2 > 0$.*

*Proof.* For the first inequality, $\Phi_{(k)}(x) \geq \frac{k}{n} \sum_{i=1}^{n} |x|_{[i]} = \frac{k}{n} l_1(x)$. Therefore, $\mathrm{M}_{\Phi_{(k)}} \geq \frac{k}{n} \mathrm{M}_{l_1} \geq \lambda_1 k/\sqrt{n}$ for some absolute constant $\lambda_1 > 0$. For the second inequality, $\Phi_{(k)}(x) \leq k l_\infty(x)$, and thus $\mathrm{M}_{\Phi_{(k)}} \leq k \mathrm{M}_{l_\infty} \leq \frac{\lambda_2 k \sqrt{\log n}}{\sqrt{n}}$ for some absolute constant $\lambda_2 > 0$. $\square$

*Proof of Theorem 6.1.* To bound $\mathrm{mmc}(\Phi_{(k)})$, consider first $n' \geq k$, then by a direct calculation,

$$\frac{\sqrt{k}}{(\lambda_2 k \sqrt{\log n'}/\sqrt{n'})} \leq \frac{\mathfrak{b}_{\Phi_{(k)}^{(n')}}}{\mathrm{M}_{\Phi_{(k)}^{(n')}}} \leq \frac{\sqrt{k}}{(\lambda_1 k/\sqrt{n'})}.$$

For $n' \leq k$, we have $\Phi_{(k)}(x) = l_1(x)$ for all $x \in \mathbb{R}^{n'}$, and we know that $\mathrm{mmc}(l_1)$ is a constant. The first part of the theorem follows.

To bound $\mathrm{mmc}(\Phi'_{(k)})$, consider first the case $n' \geq k$. For all $x \in \mathbb{R}^{n'}$ we have $\Phi'_{(k)}(x) \geq l_1(x)/k$, thus $\mathrm{M}_{\Phi'^{(n')}_{(k)}} = \Omega(\sqrt{n'}/k)$. In addition, $\mathfrak{b}_{\Phi'^{(n')}_{(k)}} \leq \max\{1, \sqrt{n'}/k\}$, and thus $\mathfrak{b}_{\Phi'^{(n')}_{(k)}} / \mathrm{M}_{\Phi'^{(n')}_{(k)}} \leq \sqrt{k}$. Consider now the case $n' \leq k$. For all $x \in \mathbb{R}^{n'}$, we have $\Phi'_{(k)}(x) = l_\infty(x)$, and thus, $\mathfrak{b}_{\Phi'^{(n')}_{(k)}} / \mathrm{M}_{\Phi'^{(n')}_{(k)}} = \Theta(\sqrt{n'/\log n'})$. We conclude that $\mathrm{mmc}(\Phi'_{(k)}) = \Omega(\sqrt{k/(\log k + 1)})$. $\square$

## 6.2 $Q$-Norms and $Q'$-Norms

A norm $l : \mathbb{R}^n \to \mathbb{R}$ is called a *$Q$-norm* if there exists a symmetric norm $\Phi : \mathbb{R}^n \to \mathbb{R}$ such that

$$\forall x \in \mathbb{R}^n, \qquad l(x) = \Phi(x^2)^{1/2},$$

where $x^p = (x_1^p, x_2^p, \ldots, x_n^p)$ denotes coordinate-wise $p$-th power. A norm $l' : \mathbb{R}^n \to \mathbb{R}$ is called a *$Q'$-norm* if its dual norm, which is given by $l(x) = \sup\{\frac{\langle x, y \rangle}{l'(y)} : y \neq 0\}$, is a $Q$-norm.

We can show that every $Q'$-norm can be approximated using polylogarithmic space, by bounding $\mathfrak{b}_{l'} / \mathrm{M}_{l'}$ and then applying Theorem 1.1, as follows.

**Theorem 6.4.** *For every $Q'$-norm $l' : \mathbb{R}^n \to \mathbb{R}$, $\mathrm{mmc}(l') = O(\log n)$.*

**Corollary 6.5** (Streaming Complexity of $Q'$-Norms). *Every $Q'$-norm $l' : \mathbb{R}^n \to \mathbb{R}$ can be $(1+\epsilon)$-approximated by a one-pass streaming algorithm that uses $\mathrm{poly}(\log(n)/\epsilon)$ space.*

The proof of Theorem 6.4 will follow by establishing the four lemmas below. It builds on the machinery developed in Section 3 to compare the median of $l$ to $l(\xi^{(n')})$, where $\xi^{(n')}$ is the $l_2$-normalized all-ones vector of dimension $n'$.

**Lemma 6.6.** *Let $l : \mathbb{R}^n \to \mathbb{R}$ be a $Q$-norm, and let $0 < n' \leq n$. Then $l(\xi^{(n')}) \geq l(\xi^{(n)})/2$.*

*Proof.* Write $n = qn' + r$, where $r < n'$ is the remainder. Then
$$\xi^{(n)} = \left( \sqrt{\tfrac{n'}{n}} (\underbrace{\xi^{(n')}, \ldots, \xi^{(n')}}_{q \text{ times}}), \sqrt{\tfrac{r}{n}} \xi^{(r)} \right).$$
By monotonicity of symmetric norms and the triangle inequality,
$$l(\xi^{(n)}) \leq l\left( \sqrt{\tfrac{n'}{n}} (\underbrace{\xi^{(n')}, \ldots, \xi^{(n')}}_{q \text{ times}}) \right) + l\left( \sqrt{\tfrac{r}{n}} \xi^{(r)} \right) \leq 2l\left( \sqrt{\tfrac{n'}{n}} (\underbrace{\xi^{(n')}, \ldots, \xi^{(n')}}_{q \text{ times}}) \right).$$
We can write $l(x) = \Phi(x^2)^{1/2}$ for some symmetric norm $\Phi$. Thus, by the triangle inequality,
$$l\left( \sqrt{\tfrac{n'}{n}} (\underbrace{\xi^{(n')}, \ldots, \xi^{(n')}}_{q \text{ times}}) \right) = \sqrt{\tfrac{n'}{n}}\, \Phi\left( (\xi^{(n')})^2, \ldots, (\xi^{(n')})^2 \right)^{1/2} \leq \sqrt{\tfrac{n'}{n}} \left( q\, \Phi((\xi^{(n')})^2) \right)^{1/2} \leq l(\xi^{(n')}).$$
□

**Lemma 6.7.** *Let $l : \mathbb{R}^n \to \mathbb{R}$ be a $Q$-norm, then for all $x \in \mathbb{R}^n$, $l(x) \leq l_2(x)$.*

*Proof.* Let $l(x) = \Phi(x^2)^{1/2}$ for some symmetric norm $\Phi$. By Lemma 2.2, $\Phi(x) \leq l_1(x)$ and therefore $l(x) = (\Phi(x^2))^{1/2} \leq l_1(x^2)^{1/2} = l_2(x)$. □

The next lemma can be viewed as a complement of Lemma 3.14 (monotonicity of the median) for the special case of $Q$-norms.

**Lemma 6.8.** *Let $l : \mathbb{R}^n \to \mathbb{R}$ be a $Q$-norm, and let $0 < n' \leq n$ be an integer. Then*
$$\mathrm{M}_{l^{(n)}} \leq \lambda \sqrt{\log n}\, \mathrm{M}_{l^{(n')}}$$
*for some absolute constant $\lambda > 0$.*

*Proof.* By Lemmas 3.12 and 6.6, we can find absolute constant $\lambda_1, \lambda_2 > 0$ such that
$$\lambda_1 \mathrm{M}_{l^{(n)}} / \sqrt{\log n} \leq l(\xi^{(n)}) \leq 2l(\xi^{(n')}) \leq 2\lambda_2 \mathrm{M}_{l^{(n')}}.$$
□

Now we show that a $Q$-norm achieves roughly the minimum at $\xi^{(n)}$.

**Lemma 6.9** (Flat Minimum). *Let $l : \mathbb{R}^n \to \mathbb{R}$ be a $Q$-norm. Then*
$$\forall x \in S^{n-1}, \qquad l(\xi^n) \leq 6\sqrt{\log n}\, l(x).$$

*Proof.* Set $\alpha := 1/2$ and fix a vector $x \in S^{n-1}$. We permute its coordinates and write $|x| = (V_1; V_2; \ldots; V_t; V')$, where $V_i = \{|x_j| : \alpha^j < |x_j| \leq \alpha^{j-1}\}$ for $i = 1, \ldots, t = \log n$, and $V' = \{|x_j| : |x_j| \leq 1/n\}$. Let $b_i = |V_i|$. Since $l_2(x) = 1$,
$$1 = l_2(x)^2 \leq \sum_{i=1}^t b_i \alpha^{2(i-1)} + 1/n.$$
Thus, there exists $i \leq t$ for which $|V_i| \alpha^{2(i-1)} \geq \tfrac{1}{2t}$, and together with Lemma 6.6,
$$l(x) \geq l(V_i) \geq \sqrt{b_i}\, \alpha^i\, l(\xi^{(b_i)}) \geq \sqrt{\tfrac{\alpha^2}{2t}}\, l(\xi^{(b_i)}) \geq \sqrt{\tfrac{1}{8t}}\, l(\xi^{(n)})/2.$$
□

*Proof of Theorem 6.4.* Let $l$ be the $Q$-norm which is dual to $l'$. By Lemma 6.9, $\forall x \in \mathbb{R}^n$, $l_2(x) \leq 6\sqrt{\log n}/l(\xi^{(n)}) \cdot l(x)$, which implies, using Fact 2.3, that $\mathfrak{b}_{l'} \leq 6\sqrt{\log n}/l(\xi^{(n)})$. By Fact 2.4 and Lemma 3.12, we know that $1/\mathrm{M}_{l'} \leq \mathrm{M}_l \leq l(\xi^{(n)})\sqrt{\log n}/\lambda_1$. The theorem follows by putting the two bounds together. □

# 7 Concluding Remarks

There is obviously a poly($\frac{1}{\epsilon} \log n$) gap between our upper and lower bounds. For the $l_p$ norms, $p > 2$, our lower bound is $\Omega(n^{1-2/p})$, matching the true space complexity to within a $\Theta(\log n)$ factor [LW13, Gan15]. Despite the gap, we do partially answer Open Problem 30 (Universal Sketching) in [sub06], by showing that the class of symmetric norms admits universal sketches, and also Open Problem 5 (Sketchable Distances) in [sub06], by showing that every symmetric norm $l$ admits streaming algorithms and is thus sketchable with space $\text{mmc}(l)^2 \cdot \text{poly}(\frac{1}{\epsilon} \log n)$.

Both our algorithm and our lower bound rely on the symmetry of the norm. It would be very interesting to see whether the modulus of concentration is a key factor in the space complexity also for general norms. Our results do extend a little towards more general norms. Notice that, given any symmetric norm $l$ on $\mathbb{R}^n$ and invertible linear transformation $A : \mathbb{R}^n \to \mathbb{R}^n$, our results also apply to the streaming complexity of $l_A := l(A(\cdot))$, which is always a norm but is generally not symmetric. For example, $l_2(A(\cdot))$ is the norm induced by the inner product $\langle x, y \rangle_A := y^T A^T A x$, and it is not symmetric unless all singular values of $A$ are the same. To compute $l_A(v)$ one applies $A$ to the incoming stream vector $v$ and then runs an algorithm for $l$ (we do not count the storage for $A$). Therefore, the space complexity of $l_A$ is no worse than that of $l$, and, as the same argument applies to $l = l_A(A^{-1}(\cdot))$, the two must have the same streaming complexity (we assume that $O(\log n)$ bits suffice to represent any entry of $A$ or $A^{-1}$ to sufficient precision). More generally, norms that can be related to each other by composition with an invertible linear transformation, as above, must have the same space complexity. On the other hand, this operation does not preserve $\text{mc}(l)$ or $\text{mmc}(l)$. Perhaps a norm should be put into a "canonical form" that is more amenable to determining its space complexity. For example, the distorted Euclidean norm $l(v) = (v^T A^T A v)^{1/2}$, mentioned above, may have $\text{mmc}(l)$ on the order of $\min\{\sqrt{n}, \sigma_1(A)/\sigma_n(A)\}$, but it can be seen immediately to have space complexity $\text{poly}(\frac{1}{\epsilon} \log n) \cdot \text{mmc}(l_2)^2$ bits (in fact $O(\frac{1}{\epsilon^2} \log n)$ bits), from the AMS algorithm [AMS99] and by recognizing $l(v) = l_2(Av)$ (again assuming $O(\log n)$ bits represents any entry of $A$ to sufficient precision). Can we use $\text{mmc}(\cdot)$ to determine the space complexity of every norm?

It would be very interesting also to design a small sketch that is oblivious to the linear transformation. For instance, let $\mathcal{M}$ be a family of linear transformations where $l_A \neq l_B$ for all $A, B \in \mathcal{M}$. Is there a linear sketch that approximates the norm $l_A(v)$ for any streamed vector $v \in \mathbb{R}^n$ and linear transformation $A \in \mathcal{M}$? Observe that no small sketch can be oblivious to all linear transformations, since that would allow recovery of every coordinate of $v$.

Our Theorem 1.3 shows a quadratic space-approximation tradeoff for every symmetric norm. Previously, this was only known for the $l_\infty$ norm due to Saks and Sun [SS02]. Investigating space-approxmation tradeoff is an interesting direction because such tradeoffs appear in the sketching lower bounds of [AKR15], however no matching algorithms are known for other specific norms of interests, such as the Earth Mover Distance and the trace norm (of matrices).

# References

[AFS12]  A. Argyriou, R. Foygel, and N. Srebro. Sparse prediction with the $k$-support norm. In *Advances in Neural Information Processing Systems 25*, pages 1457–1465. Curran Associates, Inc., 2012. Available from: http://papers.nips.cc/paper/4537-sparse-prediction-with-the-k-support-norm.pdf.

[AKO11]  A. Andoni, R. Krauthgamer, and K. Onak. Streaming algorithms via precision sampling. In *IEEE 52nd Annual Symposium on Foundations of Computer Science*, pages 363–372. IEEE Computer Society, 2011. doi:10.1109/FOCS.2011.82.

[AKR15]  A. Andoni, R. Krauthgamer, and I. Razenshteyn. Sketching and embedding are equivalent for norms. In *47th Annual ACM Symposium on Theory of Computing*, pages 479–488. ACM, 2015. doi:10.1145/2746539.2746552.

[AMS99]  N. Alon, Y. Matias, and M. Szegedy. The space complexity of approximating the frequency moments. *J. Comput. Syst. Sci.*, 58(1):137–147, 1999. doi:10.1006/jcss.1997.1545.

[ANPW13] A. Andoni, H. L. Nguyen, Y. Polyanskiy, and Y. Wu. Tight lower bound for linear sketches of moments. In *40th International Conference on Automata, Languages, and Programming*, ICALP'13, pages 25–32. Springer-Verlag, 2013. `doi:10.1007/978-3-642-39206-1_3`.

[BC15] V. Braverman and S. R. Chestnut. Universal sketches for the frequency negative moments and other decreasing streaming sums. In *APPROX/RANDOM 2015*, volume 40 of *Leibniz International Proceedings in Informatics (LIPIcs)*, pages 591–605. Schloss Dagstuhl, 2015. `doi:10.4230/LIPIcs.APPROX-RANDOM.2015.591`.

[BCWY16] V. Braverman, S. R. Chestnut, D. P. Woodruff, and L. F. Yang. Streaming space complexity of nearly all functions of one variable on frequency vectors. In *Proceedings of the 35th ACM Symposium on Principles of Database Systems*, pages 261–276, New York, NY, USA, 2016. ACM. Available from: http://doi.acm.org/10.1145/2902251.2902282, `doi:10.1145/2902251.2902282`.

[Bha97] R. Bhatia. *Matrix analysis*, volume 169 of *Graduate Texts in Mathematics*. Springer-Verlag, New York, 1997. `doi:10.1007/978-1-4612-0653-8`.

[BJKS04] Z. Bar-Yossef, T. S. Jayram, R. Kumar, and D. Sivakumar. An information statistics approach to data stream and communication complexity. *J. Comput. Syst. Sci.*, 68(4):702–732, 2004. `doi:http://dx.doi.org/10.1016/j.jcss.2003.11.006`.

[BKSV14] V. Braverman, J. Katzman, C. Seidell, and G. Vorsanger. An optimal algorithm for large frequency moments using $o(n^{1-2/k})$ bits. In *APPROX/RANDOM 2014*, volume 28 of *Leibniz International Proceedings in Informatics (LIPIcs)*, pages 531–544. Schloss Dagstuhl, 2014. `doi:10.4230/LIPIcs.APPROX-RANDOM.2014.531`.

[BO10] V. Braverman and R. Ostrovsky. Zero-one frequency laws. In *42nd ACM symposium on Theory of Computing*, pages 281–290. ACM, 2010. `doi:10.1145/1806689.1806729`.

[BO13a] V. Braverman and R. Ostrovsky. Approximating large frequency moments with pick-and-drop sampling. In *APPROX/RANDOM 2013*, volume 8096 of *Lecture Notes in Computer Science*, pages 42–57. Springer, 2013. `doi:10.1007/978-3-642-40328-6_4`.

[BO13b] V. Braverman and R. Ostrovsky. Generalizing the layering method of Indyk and Woodruff: Recursive sketches for frequency-based vectors on streams. In *APPROX/RANDOM 2013*, volume 8096 of *Lecture Notes in Computer Science*, pages 58–70. Springer, 2013. `doi:10.1007/978-3-642-40328-6_5`.

[BOR15] V. Braverman, R. Ostrovsky, and A. Roytman. Zero-one laws for sliding windows and universal sketches. In *APPROX/RANDOM 2015*, volume 40 of *Leibniz International Proceedings in Informatics (LIPIcs)*, pages 573–590. Schloss Dagstuhl, 2015. `doi:10.4230/LIPIcs.APPROX-RANDOM.2015.573`.

[CCFC02] M. Charikar, K. Chen, and M. Farach-Colton. Finding frequent items in data streams. In *Automata, Languages and Programming*, pages 693–703. Springer, 2002.

[CCM07] A. Chakrabarti, G. Cormode, and A. McGregor. A near-optimal algorithm for computing the entropy of a stream. In *18th Annual ACM-SIAM Symposium on Discrete Algorithms*, pages 328–335. SIAM, 2007.

[CDM06] A. Chakrabarti, K. Do Ba, and S. Muthukrishnan. Estimating entropy and entropy norm on data streams. *Internet Mathematics*, 3(1):63–78, 2006. `doi:10.1080/15427951.2006.10129117`.

[CKS03] A. Chakrabarti, S. Khot, and X. Sun. Near-optimal lower bounds on the multi-party communication complexity of set disjointness. In *18th IEEE Annual Conference on Computational Complexity*, pages 107–117, 2003. `doi:10.1109/CCC.2003.1214414`.

[Gan15]   S. Ganguly. Taylor polynomial estimator for estimating frequency moments. In *Automata, Languages, and Programming*, volume 9134 of *Lecture Notes in Computer Science*, pages 542–553. Springer, 2015. `doi:10.1007/978-3-662-47672-7_44`.

[GC07]    S. Ganguly and G. Cormode. On estimating frequency moments of data streams. In *APPROX/RANDOM 2007*, pages 479–493. Springer-Verlag, 2007. `doi:10.1007/978-3-540-74208-1_35`.

[Gro09]   A. Gronemeier. Asymptotically optimal lower bounds on the NIH-multi-party information complexity of the AND-function and disjointness. In *26th International Symposium on Theoretical Aspects of Computer Science STACS 2009*, pages 505–516. IBFI Schloss Dagstuhl, 2009.

[HNO08]   N. J. Harvey, J. Nelson, and K. Onak. Sketching and streaming entropy via approximation theory. In *49th Annual IEEE Symposium on Foundations of Computer Science*, pages 489–498. IEEE, 2008. `doi:10.1109/FOCS.2008.76`.

[Ind06]   P. Indyk. Stable distributions, pseudorandom generators, embeddings, and data stream computation. *J. ACM*, 53(3):307–323, 2006. `doi:10.1145/1147954.1147955`.

[IW05]    P. Indyk and D. Woodruff. Optimal approximations of the frequency moments of data streams. In *37th Annual ACM Symposium on Theory of Computing*, pages 202–208. ACM, 2005. `doi:10.1145/1060590.1060621`.

[Jay13]   T. Jayram. On the information complexity of cascaded norms with small domains. In *Information Theory Workshop (ITW), 2013 IEEE*, pages 1–5. IEEE, 2013. `doi:10.1109/ITW.2013.6691324`.

[Joh48]   F. John. Extremum problems with inequalities as subsidiary conditions. In *Studies and Essays Presented to R. Courant on his 60th Birthday*, pages 187–204. Interscience Publishers, Inc., 1948. `doi:10.1007/978-3-0348-0439-4_9`.

[JW09]    T. Jayram and D. P. Woodruff. The data stream space complexity of cascaded norms. In *50th Annual IEEE Symposium on Foundations of Computer Science*, pages 765–774. IEEE, 2009.

[KNW10]   D. M. Kane, J. Nelson, and D. P. Woodruff. On the exact space complexity of sketching and streaming small norms. In *21st Annual ACM-SIAM Symposium on Discrete Algorithms*, pages 1161–1178. SIAM, 2010.

[KV07]    B. Klartag and R. Vershynin. Small ball probability and Dvoretzky's theorem. *Israel Journal of Mathematics*, 157(1):193–207, 2007.

[Li08]    P. Li. Estimators and tail bounds for dimension reduction in $l_\alpha$ ($0 < \alpha \leq 2$) using stable random projections. In *19th Annual ACM-SIAM Symposium on Discrete Algorithms*, SODA '08, pages 10–19. SIAM, 2008. Available from: `http://dl.acm.org/citation.cfm?id=1347082.1347084`.

[LM00]    B. Laurent and P. Massart. Adaptive estimation of a quadratic functional by model selection. *Annals of Statistics*, 28(5):1302–¡96¿1338, 2000. `doi:10.1214/aos/1015957395`.

[LNW14]   Y. Li, H. L. Nguyen, and D. P. Woodruff. Turnstile streaming algorithms might as well be linear sketches. In *Proceedings of the 46th Annual ACM Symposium on Theory of Computing*, pages 174–183. ACM, 2014.

[LT13]    M. Ledoux and M. Talagrand. *Probability in Banach Spaces: isoperimetry and processes*, volume 23. Springer Science & Business Media, 2013.

[LW13]    Y. Li and D. P. Woodruff. A tight lower bound for high frequency moment estimation with small error. In *Approximation, Randomization, and Combinatorial Optimization. Algorithms and Techniques*, pages 623–638. Springer, 2013.

[MPS14] A. M. McDonald, M. Pontil, and D. Stamos. Spectral $k$-support norm regularization. In *Advances in Neural Information Processing Systems 27*, pages 3644–3652. Curran Associates, Inc., 2014. Available from: `http://papers.nips.cc/paper/5297-spectral-k-support-norm-regularization.pdf`.

[MS86] V. D. Milman and G. Schechtman. *Asymptotic theory of finite-dimensional normed spaces*, volume 1200 of *Lecture Notes in Mathematics*. Springer-Verlag, 1986. `doi:10.1007/978-3-540-38822-7`.

[Mut05] S. Muthukrishnan. Data streams: Algorithms and applications. *Found. Trends Theor. Comput. Sci.*, 1(2):117–236, 2005. `doi:10.1561/0400000002`.

[Nis92] N. Nisan. Pseudorandom generators for space-bounded computation. *Combinatorica*, 12(4):449–461, 1992.

[SS02] M. Saks and X. Sun. Space lower bounds for distance approximation in the data stream model. In *34th Annual ACM Symposium on Theory of Computing*, pages 360–369, 2002. `doi:10.1145/509907.509963`.

[sub06] List of open problems in sublinear algorithms. `http://sublinear.info/`, 2006.

[WDST14] B. Wu, C. Ding, D. Sun, and K.-C. Toh. On the Moreau–Yosida regularization of the vector $k$-norm related functions. *SIAM Journal on Optimization*, 24(2):766–794, 2014. `doi:10.1137/110827144`.

# A The Level Algorithm

In this section we prove Theorem 3.2 by presenting the level algorithm and analyzing its performance. The algorithm follows the ideas used by Indyk and Woodruff [IW05]. for approximating frequency moments. As their paper focuses on the specific problem of $l_p$-norms, its analysis is more specialized, with parameters chosen based on properties of $l_p$-norms, and it is not immediate to see how to generalize/modify it to approximate all symmetric norms. For an easier statement of our upper bound, and also for completeness, we modify their algorithm to output level vectors (instead of the value of $l_p$ norm) and repeat the analysis accordingly.

To present the high level idea, we first present a two-pass algorithm, and then modify it to be a one-pass algorithm. The two-pass algorithm is shown in Algorithm 3. Note that we assume full randomness, i.e., the algorithm has unlimited access to random bits. We can reduce the number of bits needed to $O(\log n)$ by following Indyk's method [Ind06], which uses Nisan's pseudorandom generator [Nis92], and this impacts the space complexity by an $O(\log n)$ factor.

## A.1 Two-Pass Algorithm

The purpose of this section is to prove the following theorem.

**Theorem A.1.** *There is a streaming two pass algorithm* `TwoPassLevelCounts`, *that given input stream $\mathcal{S}$ with frequency vector $v$, level base $\alpha > 1$, importance $\beta > 0$, precision $\epsilon > 0$ and error probability $\delta$, output a list $(\hat{b}_1, \hat{b}_2, \ldots, \hat{b}_t)$, where $t = \log n / \log \alpha$, such that,*

- *for all $i \in [t]$, $\hat{b}_i \leq b_i$;*
- *if $i$ is a $\beta$-important level, then $\hat{b}_i \geq (1-\epsilon)b_i$,*

*with probability at least $1 - \delta$, using space $O\left(\frac{\log^6 n}{\beta \epsilon^4} \frac{\log^2(1/\delta)}{\log \alpha} \log \frac{n^2}{\delta}\right)$.*

The high level idea of the two pass algorithm is as follows. For each $\phi = 0, 1, \ldots, O(\log n)$, we select a random subset of $[n]$ where each item is included independently with probability $2^{-\phi}$. This gives us $O(\log n)$ substreams, where each is defined by restricting the original stream to only include updates from one of the random subsets. What we will prove is that if level $i$ is important, then a random sample of the items in level $i$ will appear as heavy hitters in one of the substreams. Hence, we can find them with a CountSketch [CCFC02]. The entire sketch, that is the subsampling combined with CountSketch, is presented in Algorithm 2, SampleLevel. Finding the largest $\phi$ such that some item from level $i$ appears in the substream gives us the estimate $2^\phi$ for the size of that level, but this estimate will not be accurate enough for our purposes.

To get an accurate estimate of the level sizes we repeat the above procedure to identify heavy hitters $R = O(\epsilon^{-2} \log^2 n)$ times independently in parallel. Next, on the second pass, we determine the exact frequency of each of the items identified during the first pass and thus correctly identify its level. The final estimate for the size of level $i$ is derived by considering the largest $\phi$ such that at least a $\Omega(1/\log n)$ fraction among the repetitions with sampling probability $2^{-\phi}$ contained an item in level $i$. The entire level vector approximation procedure is Algorithm 3.

---

**Algorithm 2** SampleLevel$(\mathcal{S}, n, \beta, \epsilon, \delta, \Phi, R)$, sketch of the frequency vector by subsampling and finding $\beta$ heavy hitters

1: **Input:** stream $\mathcal{S}$, $\beta > 0$, $\epsilon > 0$, $\delta > 0$, $\Phi > 0$, $R$.
2: **Output:** Collection of maps $\{D_\phi^r \mid \phi \in [0, \Phi], r \in [0, R]\}$.
3: For each $r \in [0, R]$ and $\phi \in [0, \Phi]$ generate a substream $\mathcal{S}_\phi^r$ by sampling each $i \in [n]$ with probability $p_\phi = 2^{-\phi}$, independently, and including all updates to $i$
4: Let $D_\phi^r = \text{CountSketch}(\mathcal{S}_\phi^r, \epsilon, \beta, \delta)$.
5: Return $\{D_\phi^r \mid \phi \in [0, \Phi], r \in [0, R]\}$.

---

**Algorithm 3** TwoPassLevelCounts$(\mathcal{S}, n, \alpha, \beta, \epsilon, \delta)$, a two-pass algorithm for approximating the level vector

1: **Input:** stream $\mathcal{S}$, $\alpha > 1$, $\beta > 0$, $\epsilon > 0$ and $\delta > 0$
2: **Output:** $(\hat{b}_0, \hat{b}_1, \hat{b}_2, \hat{b}_3 \ldots \hat{b}_t)$
3: **Initialization:** Let $\Phi \leftarrow \log n$, $R \leftarrow \Theta\left(\frac{\log(1/\delta) \log^2 n}{\epsilon^2}\right)$, $\epsilon' \leftarrow \Theta(\epsilon)$
4: **First Pass:**
5: $\tilde{\mathcal{T}} \leftarrow \text{SampleLevel}(\mathcal{S}, n, O\left(\frac{\beta}{t \log(1/\delta)}\right), \epsilon', \delta/n, \Phi, R)$
6: **Second Pass:**
7: $\mathcal{T} \leftarrow$ the exact frequencies of all maps in $\tilde{\mathcal{T}}$
8: **Estimation Stage:**
9: For each $\phi \in [\Phi]$ and each $i \in [t]$, let $A_{\phi, i} \leftarrow |\{r \mid \exists i \in D_\phi^r, \alpha^{i-1} < |D_\phi^r[i]| \leq \alpha^i\}|$
10: For each $i \in [t]$, set $q_i \leftarrow \max_{\phi \in [0, \Phi]}\{\phi \mid A_{\phi, i} \geq \frac{R \log \frac{1}{\delta}}{100 \log n}\}$
11: If $q_i$ does not exist then $\hat{\eta}_i \leftarrow 0$, else $\hat{\eta}_i \leftarrow A_{q_i, i}/(R(1 + \epsilon'))$
12: If $\hat{\eta}_i = 0$ then $\hat{b}_i \leftarrow 0$, else $\hat{b}_i \leftarrow \frac{\log(1-\hat{\eta}_i)}{\log(1-2^{-q_i})}$

---

**Heavy hitter algorithm:** For convenience, we define a *map* data structure, $D$, as a set of pairs from $[n] \times \mathbb{Z}$ with the property that for each $i \in [n]$ there is at most one pair $(i, \cdot)$ in $D$. For any $i \in [n]$, we say $i \in D$ if there is a pair $(i, z)$ in $D$ and when $i \in D$ we denote $D[i] = z$ as the value paired with $i$.

Define $F_2^{\geq k}(\mathcal{S}) := \sum_{i=k+1}^{n} |v_{[i]}|^2$, where $|v_{[0]}| \geq |v_{[1]}| \ldots \geq |v_{[n]}|$ are the coordinates of the frequency vector in decreasing order with ties broken arbitrarily.

**Definition A.2.** *We call a map $D$ a $(\beta, \epsilon)$-cover of the steam $\mathcal{S}$, if*

- *if for some $i \in [n]$ such that $|v_i|^2 \geq \beta F_2^{\geq \lfloor 1/\beta \rfloor}(\mathcal{S})$ then $i \in D$;*
- *for every $j \in D$, $|v_j| \leq D[j] \leq (1 + \epsilon)|v_j|$;*

*We omit $\mathcal{S}$ if it is clear from context.*

The purpose of Algorithm 2 is to find a $(\beta, \epsilon)$-cover for each of $O(\log n)$ substreams. We call $\epsilon$ the *precision parameter*, $\beta$ the *heaviness parameter*, and $\delta$ the *error rate*[2].

---

[2]We asume $\delta < \epsilon$ for the following analysis.

**Theorem A.3** ([CCFC02]). *There is a one pass streaming algorithm* `CountSketch`$(\mathcal{S}, \epsilon, \beta, \delta)$ *that, for any input stream $\mathcal{S}$ of universe $[n]$, with frequency vector $v = (v_1, v_2, \ldots, v_n)$, outputs a map $D$ such that, with probability at least $(1 - \delta)$,*

- *$D$ is a $(\beta, \epsilon)$-cover of $\mathcal{S}$ and*
- *$|D| \leq 2/\beta$.*

*CountSketch uses $O(\frac{1}{\beta \epsilon^2} \log \frac{n}{\delta} \log n)$ bits of space.*

Consider the algorithm `SampleLevel`, we define $\mathcal{E}_\phi^r$ as the event that the $(\phi, r)$ instance of `CountSketch` outputs a $(\beta, \epsilon)$-covers for the substream $\mathcal{S}_\phi^r$. By Theorem A.3, we have $\Pr[\mathcal{E}_\phi^r] \geq 1 - \frac{\delta}{n}$. Note that $\mathcal{E}_\phi^r$ is independent of $\mathcal{S}_\phi^r$. To simplify analysis, we will assume that the output of each `CountSketch` is correct. Our choice of the error rate $\delta$ will be such that this happens with probability near 1. For the following analysis, we consider Algorithm `TwoPassLevelCounts` and assume $\epsilon = 1/\operatorname{polylog} n, \delta = 1/\operatorname{poly}(n), \beta = 1/\operatorname{polylog} n$.

**Detectability of Levels:**

**Definition A.4.** *A level $i$ of stream $\mathcal{S}$ is $\beta$-detectable, if*
$$\alpha^{2i} \geq \beta F_2^{\geq \lfloor 1/\beta \rfloor}(\mathcal{S}).$$
*We omit $\mathcal{S}$ if it is clear from the context.*

If a level $i$ is $\beta$-detectable then a `CountSketch`, with heaviness $\beta$, will include $B_i$ in its output. The following lemma is about the detectability of an important level.

**Lemma A.5.** *Suppose substream $\mathcal{S}'$ is obtained by subsampling the original stream with probability $p$. Assume $\alpha \leq 2$. If $i$ is a $\beta$-important level, then for any $\lambda > t$, with probability at least $1 - t \exp\left[-\frac{\lambda p b_i}{t \beta}\right]$, level $i$ is $\frac{\beta}{\lambda p b_i}$-detectable. In particular, if $pb_i = O(1)$, then with probability at least $1 - t \exp\left[-\Omega\left(\frac{\lambda}{t \beta}\right)\right]$, level $i$ is $\frac{\beta}{\lambda}$-detectable.*

*Proof.* Let $V'$ be the new frequency vector the substream. Let $(b'_0, b'_1, \ldots)$ be the new level sizes. Thus, for all $j \in [t]$,
$$E(b'_j) = pb_j. \tag{10}$$
By the definition of important level
$$E[b'_i] \geq \beta E[\sum_{j > i} b'_j]$$
and
$$E[b'_i \alpha^{2i}] \geq \beta E[\sum_{j \leq i} b'_j \alpha^{2j}].$$

To have level $i$ detectable, it has to be among the top $\lambda pb_i / \beta$ elements, thus $\sum_{j > i} b'_j \leq \lambda pb_i / \beta$. The probability this does not happen is, by Chernoff's bound,
$$\Pr[\sum_{j > i} b_j > \frac{\lambda pb_i}{\beta}] \leq \exp[-\Omega(\frac{\lambda pb_i}{\beta})].$$
On the otherhand, for level $i$ to be detectable $\alpha^{2i} \geq \frac{\beta}{\lambda pb_i} \sum_{j \leq i} b'_j \alpha^{2j}$. Thus, the complement occurs with probability,
$$\Pr\left[\frac{\beta}{\lambda pb_i} \sum_{j \leq i} b'_j \alpha^{2j} > \alpha^{2i}\right] \leq \Pr\left[\exists j, \quad b'_j \alpha^{2j} \geq \frac{\lambda pb_i \alpha^{2i}}{t\beta}\right]$$
$$\leq \sum_j \Pr[b'_j \alpha^{2j} \geq \frac{\lambda pb_i \alpha^{2i}}{t\beta}].$$
By Chernoff's bound and because $E(b'_j \alpha^{2j}) \leq pb_i \alpha^{2i}$,
$$\Pr\left[\frac{\beta}{\lambda pb_i} \sum_{j \leq i} b'_j \alpha^{2j} > \alpha^{2i}\right] \leq t \exp\left[-\Omega\left(\frac{\lambda pb_i \alpha^{2i}}{\alpha^{2j} t \beta}\right)\right] \leq t \exp\left[-\Omega\left(\frac{\lambda pb_i}{t\beta}\right)\right]. \qquad \square$$

**Subsample the important levels:** Let us define $\eta_{i,\phi} := 1 - (1 - p_\phi)^{b_i}$, which is the probability that at least one element from $B_i$ is sampled when the sampling probability is $p_\phi = 2^{-\phi}$. Set $\lambda = \Theta(t \log \frac{1}{\delta})$. Let $\eta'_{i,\phi}$ be the probability that an element from $B_i$ is contained in $D^1_\phi$, which is the same as the probability that it is contained in $D^r_\phi$ for any other $r$.

**Lemma A.6.** *For any level $i$ we have $\eta'_{i,\phi} < \eta_{i,\phi}$. Also, assume $\delta < \epsilon$, if level $i$ is a $\beta$-important level and $p_\phi b_i = O(1)$, then $\eta'_{i,\phi} \geq (1 - \Theta(\epsilon))\eta_{i,\phi}$.*

*Proof.* For an element to be in $D^r_\phi$ it must first be sampled, thus $\eta'_{i,\phi} \leq \eta_{i,\phi}$. On the other hand, with probability at least $1 - t \exp\left[-\Omega(\frac{\lambda p b_i}{t\beta})\right] = 1 - O(\delta)$, level $i$ is $\frac{\beta}{\lambda p b_i}$-detectable. Note that $\frac{\beta}{\lambda p b_i} = \Theta(\frac{\beta}{t \log(1/\delta)})$. Thus with probability at least $\eta_{i,\phi}(1 - \Theta(\delta)) \geq \eta_{i,\phi}(1 - O(\epsilon)))$ an element from $B_i$ is sampled and the element is reported by `CountSketch`. □

We now show that $\hat{\eta}_i$, at Line 11 of Algorithm 3, is a good estimator for $\eta'_{i,q_i}$.

**Lemma A.7.** *At Line 11, with probability at least $1 - \delta^{\Omega(\log n)}$, for all $i \in [t]$, if $\hat{\eta}_i \neq 0$ then $(1 - O(\epsilon))\eta'_{i,q_i} \leq \hat{\eta}_i \leq \eta'_{i,q_i}$. The probability is taken over the probability space of all the random bits chosen for the algorithm.*

*Proof.* If $\hat{\eta}_i \neq 0$, then $A_{q_i,i} \geq \gamma = \frac{R \log \frac{1}{\delta}}{100 \log n}$. For a particular $i$, since the sampling process is independent for each $r \in [R]$, the claim is that $E(A_{q_i,i}) = R\eta_{i,q_i} = \Omega(\gamma)$. This holds because otherwise, by a Chernoff bound, $\Pr[A_{q_i,i} \geq \gamma] = o(\delta^{\Omega(\log n)})$. Therefore by a Chernoff bound,
$$\Pr[|A_{q_i,i} - R\eta'_{i,q_i}| \geq \epsilon R\eta_{i,q_i}] \leq \exp(-\Omega(\epsilon^2 \gamma)) = \delta^{\Omega(\log(n))}.$$
Since $t = \text{polylog}(n)$, by union bound, $(1 - \epsilon')\eta_{i,q_i} \leq A_{q_i,i}/R_i \leq (1 + \epsilon')\eta_{i,q_i}$ happens for all $i \in [t]$ with probability at least $1 - \delta^{\Omega(\log(n))}$. Thus the lemma is proved. □

**Lemma A.8.** *At Line 11, if level $i$ is important, then with probability at least $1 - \delta^{\Omega(\log n)}$, the maximizer $q_i$ is well defined. The probability is taken over the probability space of all the random bits chosen for the algorithm.*

*Proof.* By Lemma A.6 we know when $p_i b_i = O(1)$ level $i$ is at least $\Omega(\frac{\beta}{t \log(1/\delta)})$-detectable. On the other hand, consider the case when $p_i = 2^{-\phi_0} = \frac{1}{b_i}$, we have $\eta_{i,\phi_0} = 1 - (1 - p_i)^{b_i} \geq 1/e$. Thus $E(A_{\phi_0,i}) \geq R/e$, thus by Chernoff's bound,
$$\Pr[A_{\phi_0,i} \leq \frac{R \log 1/\delta}{100 \log n}] \leq \exp\left[-\Omega\left(\frac{R \log 1/\delta}{\log n}\right)\right] \leq \delta^{\Omega(\log n)}.$$
Thus, there exists a $q_i \geq \phi_0$ with probability at least $1 - \delta^{\Omega(\log n)}$. Since there are at most $t = \text{polylog}(n)$ important levels, with probability at least $1 - \delta^{\Omega(\log n)}$, the corresponding value of $q_i$ is well defined for all important levels. □

**From probability estimation to size of the level:** Now we show that the conversion from $\hat{\eta}_i$ to $\hat{b}_i$ gives a good approximation to $b_i$.

**Lemma A.9.** *At line 12 of Algorithm 3. Suppose $q_i \geq 1$, $\epsilon \leq 1/2$, and $n$ is sufficiently large. If $\hat{\eta}_i \leq \eta_{i,q_i}$ then $\hat{b}_i \leq b_i$. If $\hat{\eta}_i \geq (1 - \epsilon)\eta_{i,q_i}$ then $\hat{b}_i \geq (1 - O(\epsilon))b_i$.*

*Proof.* Note that
$$b_i = \frac{\log(1 - \eta_{i,q_i})}{\log(1 - 2^{-q_i})},$$
is a increasing function of $\eta_{i,q_i}$. Thus if $\hat{\eta}_i \leq \eta_{i,q_i}$, then $\hat{b}_i \leq b_i$. On the other hand, if $\hat{\eta}_i \geq (1 - O(\epsilon))\eta_{i,q_i}$ we have
$$\hat{b}_i \geq b_i + \frac{\epsilon' \eta_{i,q_i}}{(1 - \eta_{i,q_i}) \log(1 - 2^{-q_i})} \geq b_i - O(\epsilon) b_i. \qquad \square$$

**Full Proof:**

*Proof of Theorem A.1.* Define the event $\mathcal{E}_2$ that for all important levels, $q_i$ is well defined. Define event $\mathcal{E}_3$ that for all $i \in [t]$, if $\hat{\eta}_i > 0$ then $(1 - O(\epsilon))\eta'_{q_i,i} \leq \hat{\eta}_i \leq \eta'_{q_i,i}$. By Lemma A.7 and A.8, we have that $\Pr[\mathcal{E}_2 \cap \mathcal{E}_3] \geq 1 - \delta^{O(\log n)}$. When the output of every CountSketch is correct, it follows from lemmas A.6, A.7, and A.9 the algorithm outputs an approximation to the level vector meeting the two criteria.

CountSketch is used a total of $\Phi R$ times, each with error probability at most $\delta/n$. By a union bound, the failure probability is at most $(\text{polylog } n)\delta/n = o(\delta)$. Therefore, the total failure probability of the algorithm is at most $1 - o(\delta)$.

The last step of the proof is to bound the space used by the algorithm. For each instance of the CountSketch, the space used is $O(\frac{1}{\epsilon^2} \frac{t \log(1/\delta)}{\beta} \log \frac{n^2}{\delta} \log n)$. There are $\Phi R$ instances. Finally, we can reduce the number bits need to $O(\log n)$ by using Nisan's pseudorandom generator [Nis92] with Indky's method [Ind06], which impacts the storage by a $O(\log n)$ factor. To store the level vector, it requires $t$ counters. Therefore, the total space used is,

$$O\left(\frac{\log^6 n \log^2(1/\delta)}{\beta \epsilon^4 \log \alpha} \log \frac{n^2}{\delta} + \frac{\log^2 n}{\log \alpha}\right) \tag{11}$$

bits of memory. □

## A.2 One-Pass Algorithm

In this section, following [IW05]'s approach, we show that we can convert the two-pass algorithm to a one pass algorithm by randomizing the boundary. The randomizing scheme works by changing $\alpha = 1 + \gamma$ to $\alpha' = 1 + x\gamma$, where $x$ is chosen uniformly at random from $[1/2, 1]$. Note that in [IW05], they randomize the boundary by changing the level boundaries to $x\alpha^i$, we will verify that their proof still works if choose our boundary randomization scheme. Here we assume that the algorithm works for real numbers. The full approach in [IW05] proves that the real number can be represented by first few bits of the real number $x$ and yeilds no precision loss.

We refer our algorithm as Level1. The idea is to remove the second pass in Algorithm TwoPass-LevelCounts, and just use the approximated values $D[j]$s. Note that in TwoPassLevelCounts, we require the second pass to measure the exact frequencies, so that we can accurately decide which level the sampled frequency belongs to. The task of deciding levels is to test whether $|v_j| \geq \alpha^i$. However, for each returned map $D$, since $|v_j| \leq D[j] \leq (1+\epsilon)|v_j|$, if the boundary $\alpha^i \in [|v_j|, (1+\epsilon)|v_j|]$, then $D[j] \geq \alpha^i$ may not imply $|v_j| \geq \alpha^i$.

Following [IW05], after the running of SampleLevel, for a returned map $D$, for each $j \in D$, we claim level $w = \lceil \log D[j]/\log \alpha' \rceil$ is detected. Thus whether a level $w$ is detected, is determined by the *first* $j \in [m]$ that $D[j]$ falls in interval $(\alpha'^{w-1}, \alpha'^w]$. Denote $v = |v_j|$ as the actual frequency. We test whether the event $\mathcal{E}_w : \alpha'^{w-1} \geq D[j]/(1+\epsilon)$ happens. If true, we discard the returned map $D$. If this is not the case, we know that $|v| \in (\alpha'^{w-1}, \alpha'^w]$, and therefore $w$ is the correct classification of $|v|$.

First note that if $\alpha < 2$, then the case $w = 1$ is always a good case since $D[j] \in (1, \alpha'] \Rightarrow v \in (1/(1+\epsilon), \alpha']$, which is only possible when $|v| = 1$. Then we can combine the case $w = 1$ and $w = 0$. When $w > 1$, assuming the precision parameter of CountSketch is $\epsilon'$, we bound the probability of the bad cases, which imply,

$$(1-\epsilon')|v| \leq (1 + x\gamma)^{w-1} \leq (1+\epsilon')|v|$$
$$\Rightarrow \log(1 - \epsilon') + \log|v| \leq (w-1)\log(1 + x\gamma) \leq \log|v| + \log(1 + \epsilon')$$
$$\Rightarrow \frac{\log|v|}{\gamma} - \frac{\epsilon'}{\gamma} \leq (w-1)x \leq \frac{\log|v|}{\gamma} + \frac{\epsilon'}{\gamma}$$
$$\Rightarrow \frac{\log|v|}{(w-1)\gamma} - \frac{\epsilon'}{(w-1)\gamma} \leq x \leq \frac{\log|v|}{(w-1)\gamma} + \frac{\epsilon'}{(w-1)\gamma} \tag{12}$$

Since $w > 1$. Thus $\Pr[\mathcal{E}_w] \leq \frac{4\epsilon'}{(w-1)\gamma} \leq \frac{4\epsilon'}{\gamma}$. We can choose the precision parameter as, $\epsilon' = \frac{\gamma^2}{8 \log n}$. And there are at most $t' = \log n/\log \alpha' = O(\log n/\gamma)$ detected levels. $\mathcal{E}_w$ happens for any of the detected levels $w \in [t']$ with probability at most,

$$\frac{\log n}{\log \alpha'} \frac{\epsilon'}{\gamma} \leq \frac{\epsilon' \log n}{\gamma^2} = \frac{1}{2}.$$

Thus, by parallel repeating the SampleLevel algorithm $\log \frac{n^2}{\delta}$ times, with probability at list $1 - O(\delta/n^2)$,

there is a good map $D$ in every $\log \frac{n^2}{\delta}$ many returned maps. Thus, with probability at least $1 - O(\delta/n)$ we can still find as many as good maps as in the two-pass algorithm. The remaining analysis of the algorithm follows exactly as in the two pass algorithm. The memory used in the one-pass modification can be obtained by replacing one of the $1/\epsilon^2$ factor in Equation (11) with $1/\epsilon'^2$ and multiply with $\log \frac{n^2}{\delta}$,

$$O\left(\frac{\log^8 n}{\beta \epsilon^2} \frac{\log^2(1/\delta)}{\log^5 \alpha} \log^2 \frac{n^2}{\delta}\right)$$

bits. Thus we yield,

**Theorem A.10.** *There is a streaming one pass algorithm* `Level1`, *that given input stream $\mathcal{S}$ with frequency vector $v$, level base $\alpha = 1 + \gamma$, importance $\beta > 0$, precision $\epsilon > 0$ and error probability $\delta$, output a list $(\alpha' = 1 + \Theta(\gamma), \hat{b}_1, \hat{b}_2, \ldots, \hat{b}_t)$, where $t = \log n / \log \alpha'$, such that,*

- *for all $i \in [t]$, $\hat{b}_i \leq b_i$;*

- *if $i$ is a $\beta$-important level, then $\hat{b}_i \geq (1 - \epsilon) b_i$,*

*with probability at least $1 - \delta$, using space $O\left(\frac{\log^8 n}{\beta \epsilon^2} \frac{\log^2(1/\delta)}{\log^5 \alpha} \log^2 \frac{n^2}{\delta}\right)$ bits.*